\documentclass[aps,pra,twocolumn,reprint,superscriptaddress]{revtex4}

\pdfoutput=1
\usepackage{hyperref}
\usepackage{graphicx}  
\usepackage{bm}        
\usepackage{amssymb}   
\usepackage{xspace}
\usepackage{verbatim}
\usepackage{color}
\usepackage{soul}
\usepackage{subfigure}
\usepackage[export]{adjustbox}
\usepackage{dcolumn}
\usepackage{amsthm,stmaryrd,mathtools} 
\usepackage{txfonts}
\usepackage{braket}
\usepackage{amsmath}
\usepackage{xcolor}
\usepackage{array}
\usepackage{multirow}
\usepackage{tabularx}
\usepackage{booktabs}
\usepackage{lipsum}

\begin{document}


\title{Simultaneous multiple-user quantum communication across a spin-chain channel}

\author{Rozhin Yousefjani}%
\email{RozhinYousefjani@uestc.edu.cn}
\affiliation{Institute of Fundamental and Frontier Sciences, University of Electronic Science and Technology of China, Chengdu 610051, China}
\author{Abolfazl Bayat}
\email{abolfazl.bayat@uestc.edu.cn}
\affiliation{Institute of Fundamental and Frontier Sciences, University of Electronic Science and Technology of China, Chengdu 610051, China}

\begin{abstract}
The time evolution of spin chains has been extensively studied for transferring quantum states between
different registers of a quantum computer. Nonetheless, in most of these protocols only one sender-receiver
pair can share the channel at each time. This significantly limits the rate of communication in a network of
many users because they can only communicate through their common data-bus sequentially and not all at the
same time. Here, we propose a protocol in which multiple users can share a spin chain channel simultaneously
without having crosstalk between different parties. This is achieved by properly tuning the local parameters
of the Hamiltonian to mediate an effective interaction between each pair of users via a distinct set of energy
eigenstates of the system. We introduce three strategies with different levels of Hamiltonian tuning; each might
be suitable for a different physical platform. All the three strategies provide very high transmission fidelities with
vanishingly small crosstalk. The protocol is robust against various imperfections and we specifically show that
our protocol can be experimentally realized on currently available superconducting quantum simulators.     
\end{abstract}

\maketitle


\section{Introduction}  
Spin chains have been proposed~\cite{bose2003quantum} and extensively studied~\cite{bose2007quantum,nikolopoulos2014quantum} as data-bus for transferring quantum information between different registers through their natural time evolution. The main advantage of these protocols is their minimal demand for dynamical control and their resilience against disorder and imperfections~\cite{petrosyan2010state,yang2010spin}. The drawback, however, is the dispersive nature of their dynamics which scrambles the information among various degrees of freedom~\cite{eisert2015quantum,lewis2019dynamics}. Many proposals have been put forward to fix this issue. By engineering the couplings~\cite{christandl2004perfect,christandl2005perfect,di2008perfect} or tuning long range exchange interactions~\cite{kay2006perfect} one can achieve a linear dispersion relation and thus fulfill perfect state transfer. Simpler designs excite the system only in the linear zone of its dispersion relation and achieve pretty good transfer fidelities~\cite{apollaro201299,banchi2011nonperturbative,yao2011robust}. Dual rail systems~\cite{burgarth2005conclusive,burgarth2005perfect} and $d$-level spin chains~\cite{bayat2014arbitrary,qin2013high} can asymptotically reach perfect state transfer. Adiabatic attachment and detachment of qubits~\cite{chancellor2012using,farooq2015adiabatic,mohiyaddin2016transport} and their faster versions through a short cut to adiabaticity~\cite{huang2018quantum,baksic2016speeding}, optimal control~\cite{caneva2009optimal} and machine learning assisted transfer~\cite{porotti2019coherent} have also been suggested. Routing information between different nodes of a graph can be achieved by a combination of ferro and anti-ferromagnetic couplings~\cite{pemberton2011perfect,karimipour2012perfect} and encoding the information in a decoherence free subspace protects it against noise~\cite{qin2015protected}. Exploiting projective measurements for encoding~\cite{pouyandeh2014measurement} and countering dephasing~\cite{bayat2015measurement} can enhance quality of transfer and local rotations~\cite{burgarth2007optimal,yang2011entanglement} may yield an enhanced communication rate. In addition, an important class of protocols relies on inducing an effective end-to-end interaction between the sender-receiver sites through either weak boundary couplings~\cite{wojcik2005unmodulated,venuti2006long,venuti2007long,paganelli2013routing,lemonde2019quantum,lemonde2018phonon,Apollaro2019} or large magnetic fields near the ends~\cite{lorenzo2013quantum,apollaro2015many}. 
Some of the proposals have been experimentally implemented in coupled optical fibers~\cite{bellec2012faithful,perez2013coherent}, nuclear magnetic resonance devices~\cite{PhysRevA.90.012306}, optical lattices~\cite{fukuhara2013quantum} and superconducting quantum simulators~\cite{li2018perfect}.

\begin{figure}[h!]
\includegraphics[width=0.9\linewidth]{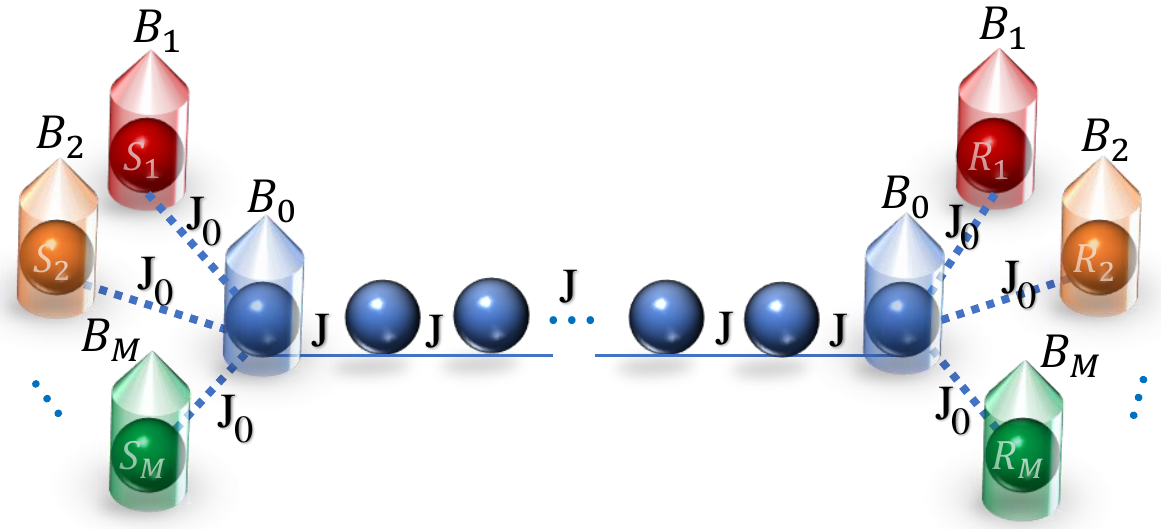}
\caption{Schematic of simultaneous quantum communication between
multiple users across a spin-chain channel. By optimizing the
local parameters at the sender and receiver sites, multiple pairs can
use the channel simultaneously. }\label{fig:Configuration}
\end{figure}

In almost all the existing state-transfer protocols only one
sender-receiver pair can use the spin-chain channel at each
time. This significantly reduces the communication rate, a bottleneck that may ultimately limit the speed of big quantum computers.
Although multiple qubit communication~\cite{apollaro2015many,Apollaro2019} have been proposed they have no freedom to adjust the choice of the sender and receiver qubits which are predetermined by the symmetry of the system and thus still work as a single sender-receiver protocol with multiple qubits.  Alternatively, to increase the rate, bi-directional protocols have been proposed but they have poor fidelities~\cite{wang2011duplex}. In classical communication networks (e.g. telecommunication systems), however, the frequency bandwidth of the channel is divided between multiple users who can use the channel simultaneously. This can be achieved by modulating the signal of each pair of sender-receivers with a different carrier signal, each with a distinct frequency, and send it through a common  channel. Since each sender's data lies in a different frequency bandwidth, the corresponding receiver can access the relevant information by using a proper frequency filter. Consequently, crosstalks are prevented and the communication rate is significantly enhanced.  A key open question is whether one can develop a quantum counterpart of classical communication systems and allow multiple users simultaneously communicating through a common channel.

In this paper, we address this critical problem by proposing
a communication scheme which is based on tuning the
local parameters at the sender and receiver sites. These local
tunings, proposed in three different strategies, excite different
sets of energy eigenstates for communication of each pair
of users and thus result in high transmission fidelities and
negligible crosstalk. We have also shown that our protocol is 
stable against various sources of imperfections and propose to
implement it on superconducting quantum simulators.


\section{The Model}
We consider $M$ sender-receiver pairs in a way that pair $\alpha$ ($\alpha=1,\cdots,M$) communicate between the qubits $S_\alpha$ (sender) and $R_\alpha$ (receiver). All pairs share a common spin chain data-bus between their sender and receiver sites. A schematic of the system is given in Fig.~\ref{fig:Configuration}. The goal is to establish simultaneous high-fidelity communication between any pair of $(S_\alpha,R_\alpha)$ while suppressing the crosstalk between $(S_\alpha,R_\beta)$ with $\alpha\neq\beta$. The spin chain channel consists of $N$ spin-$1/2$ particles which interact via Hamiltonian 
\begin{equation}\label{eq:channel Hamiltonian}
H_{ch}=J\textstyle \sum_{i=1}^{N-1}(\sigma_{i}^{x}\sigma_{i+1}^{x}+\sigma_{i}^{y}\sigma_{i+1}^{y})+B_0(\sigma_{1}^{z}+\sigma_{N}^{z}),
\end{equation}
where $\sigma_{i}^{x,y,z}$ are the Pauli operators acting on site $i$, $J$ is the spin exchange coupling and $B_0$ is the magnetic field in the $z$ direction acting only on sites $1$ and $N$. All senders (receivers) are coupled to the first (last) site of the channel. The interaction between the users' qubits and the channel is given by  
\begin{eqnarray}\label{eq:Interaction Hamiltonian}
H_I&=&J_0 \textstyle\sum_{\alpha=1}^{M}\left(\sigma_{S_\alpha}^{x}\sigma_{1}^{x}+\sigma_{S_\alpha}^{y}\sigma_{1}^{y}+\sigma_{N}^{x}\sigma_{R_\alpha}^{x}+\sigma_{N}^{y}\sigma_{R_\alpha}^{y} \right)\cr \cr
&+&
\textstyle\sum_{\alpha=1}^{M} B_{\alpha}(\sigma_{S_\alpha}^{z}+\sigma_{R_\alpha}^{z}),
\end{eqnarray}
where $J_0$ is the coupling between the users and the channel and $B_\alpha$ is the magnetic field acting on the pair user $\alpha$ (see Fig.~\ref{fig:Configuration}). 
Without loss of generality, we assume that the sender $\alpha$ initially sets its qubit in an arbitrary, possibly unknown, state 
\begin{eqnarray}
\vert\psi_{_{S_\alpha}}\rangle=\cos(\tfrac{\theta_{\alpha}}{2})\vert 0 \rangle + e^{i\phi_{\alpha}}\sin(\tfrac{\theta_{\alpha}}{2})\vert 1 \rangle,
 \quad \alpha=(1,\cdots,M)
\end{eqnarray}
where $\theta_\alpha$ and $\phi_{\alpha}$ are the angles determining the quantum state on the surface of the Bloch sphere. 
The rest of the spins, including all receivers and the channel, are initialized in $\vert 0\rangle$. Therefore, the state of the whole system becomes
\begin{eqnarray}\label{eq:Initial state}
\textstyle  \vert \Psi_{0} \rangle {=}\vert\psi_{_{S_1}}\rangle {\otimes}{\cdots}{\otimes} \vert\psi_{_{S_M}}\rangle \otimes \vert\bm{0}_{ch}\rangle \otimes \vert 0 _{_{R_1}}\rangle {\otimes} {\cdots}{\otimes} \vert 0_{_{R_M}}\rangle,
 \end{eqnarray}
where $\vert\bm{0}_{ch}\rangle{=}\vert0,\cdots,0\rangle$ shows the state of the channel.
Since this quantum state is not an eigenstate of the total Hamiltonian $H {=} H_{ch} {+} H_{I}$, it evolves as $\vert \Psi(t)\rangle {=} e^{-iHt} \vert \Psi_0\rangle$. At any time $t$ the state of the receiver sites are given by $\rho_{_{R_\alpha}}(t){=}Tr_{\widehat{R}_\alpha} \left( \vert \Psi(t)\rangle \langle \Psi(t) \vert  \right)$,  
where $Tr_{\widehat{R}_{\alpha}}$ means tracing over all sites except $R_\alpha$.  
\begin{figure}[t!]
	\centering\offinterlineskip
	\includegraphics[width=\linewidth]{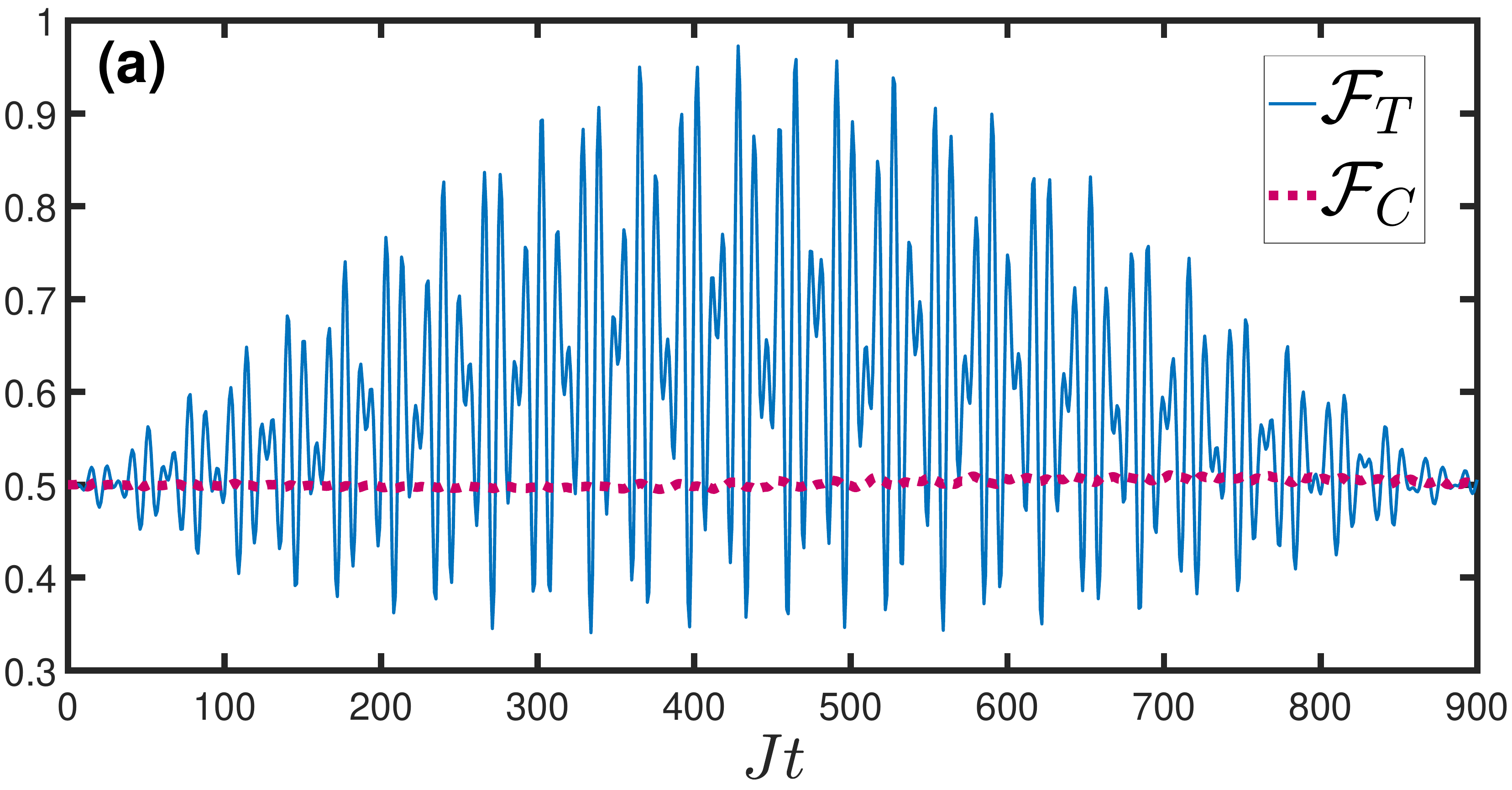}
	\includegraphics[width=\linewidth , height=2.7cm]{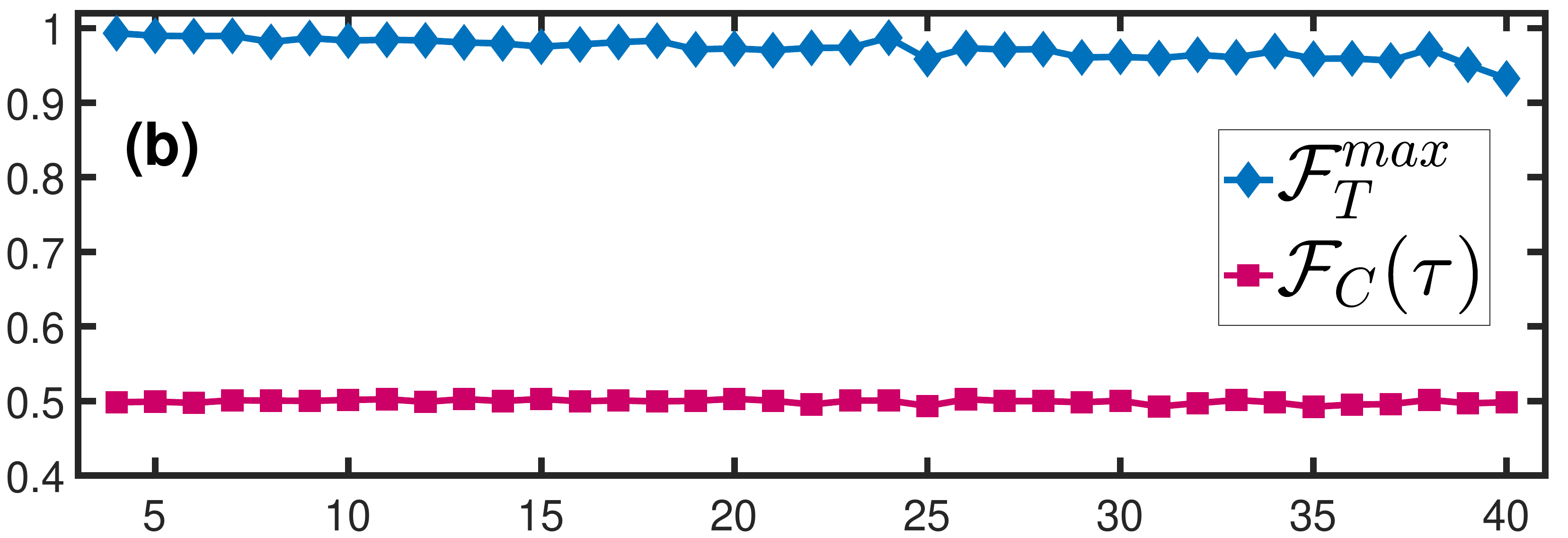}
	\includegraphics[width=\linewidth , height=2.6cm]{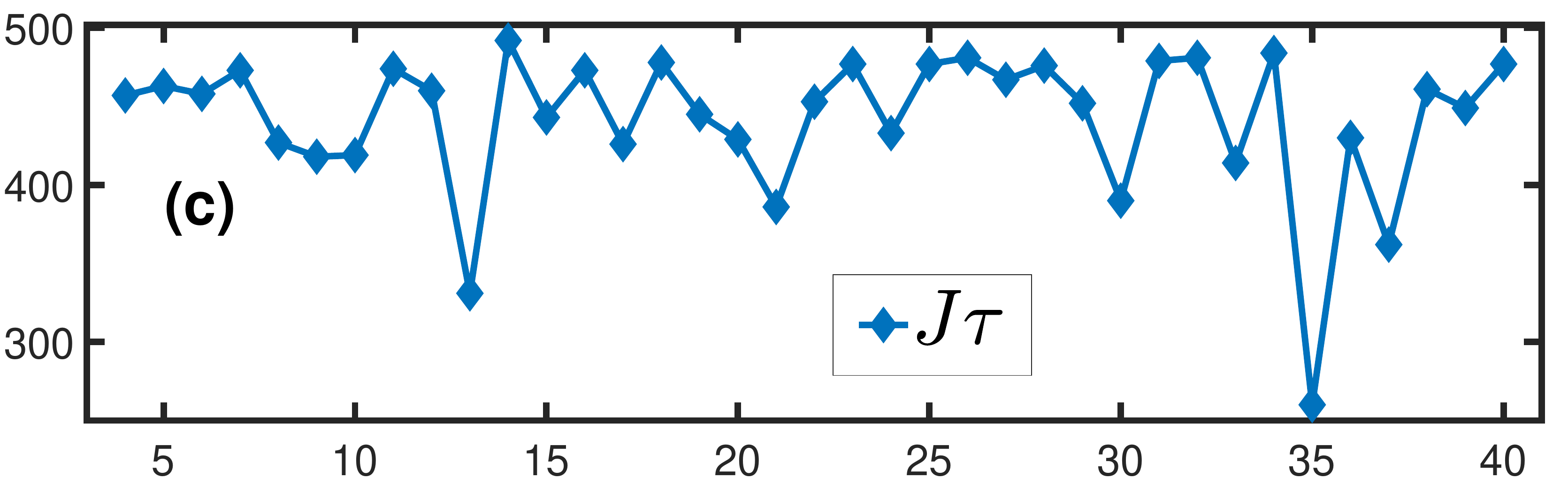}
	\includegraphics[width=\linewidth , height=2.8cm]{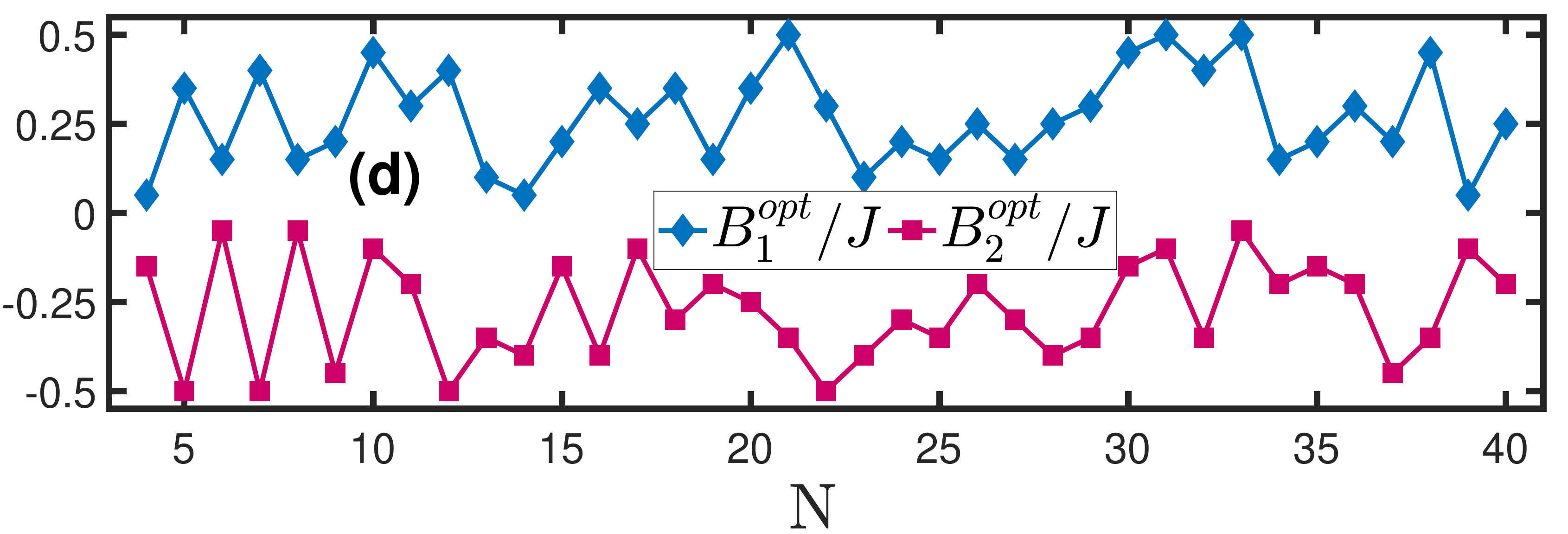}
	\caption{Strategy 1: (a) The average transmission fidelity $\mathcal{F}_T$ and $\mathcal{F}_C$ as functions of time in a spin chain of length $N{=}20$. The Hamiltonian parameters are taken to be $J_{0}/J{=}0.04$, $B_{1}/J{=}0.35$, $B_{2}/J{=}{-}0.25$. (b) The obtainable average transmission fidelity $\mathcal{F}_{T}^{max}$ and corresponding average crosstalk $\mathcal{F}_{C}(\tau)$ as functions of $N$. (c) The optimal time $\tau$, optimized over the interval $\tau{\in}[1,500]/J$, at which the average transmission fidelity peaks. (d)  The optimal parameters $B_{1}^{opt}/J{\in}[0.05,0.5]$ and $B_{2}^{opt}/J{\in}[-0.5,-0.05]$ when the the optimal coupling is found to be $J_{0}^{opt}/J{=}0.04$. All the plotted quantities are dimensionless.}\label{fig:S1}
\end{figure}
To quantify the quality of transfer between the sender $\alpha$ and receiver $\beta$ we define a fidelity matrix as $F_{\alpha\beta}(t,\Theta){=}\langle \psi_{_{S_\alpha}}\vert \rho_{_{R_\beta}}(t) \vert \psi_{_{S_\alpha}} \rangle$~($\alpha,\beta{=}1,\cdots,M$), where $\Theta{=}\{\theta_1,\cdots,\theta_M,\phi_1,\cdots, \phi_M \}$ accounts for the input parameters of the senders. 
To get an input-independent quantity one can take the average of these fidelities over all
possible initial states on the surface of the Bloch spheres for all $M$ users 
\begin{equation}
\overline{F}_{\alpha\beta}(t)=\textstyle {\int} F_{\alpha\beta}(t,\Theta) d\Omega_{1}\cdots d\Omega_{M},
\end{equation}
where $d\Omega_{\alpha}=\frac{1}{4\pi}\sin(\theta_{\alpha})d \theta_{\alpha} d \phi_{\alpha}$ is the normalized $SU{(2)}$ Haar measure.
For our Hamiltonian $H$ that conserves the total number of excitations, we provide a general form of $\overline{F}_{\alpha\beta}(t)$ in Appendix A.
The diagonal term $\overline{F}_{\alpha\alpha}(t)$ quantifies the average fidelity of the transmission between the sender-receiver $\alpha$ and the off diagonal term $\overline{F}_{\alpha\beta}(t)$ with $\alpha\neq \beta$ accounts for the crosstalk between the users $\alpha$ and $\beta$. 
Our goal is to maximize the transmission fidelities $\overline{F}_{\alpha\alpha}$ simultaneously and meanwhile keeping the crosstalk fidelities $\overline{F}_{\alpha\beta} $ around $0.5$ (i.e. no crosstalk), through controlling the Hamiltonian parameters $B_0$, $J_0$ and $B_\alpha$'s. 
This goal can be pursue by maximizing the average of the transmission fidelities $\mathcal{F}_T{=}\sum_{\alpha=1}^{M}\overline{F}_{\alpha\alpha}/M$ in time and, consequently, keeping  the average of the crosstalks $\mathcal{F}_C{=}\sum_{\alpha \neq \beta=1}^{M}\overline{F}_{\alpha\beta}/M(M-1)$ around 0.5.   
Our protocol can be understood in two steps.    
The first step is to induce an effective end-to-end transmission between the senders and the receivers, namely confining the excitations to the subspace $\{S_1,\cdots ,S_M,R_1,\cdots ,R_M\}$ and leaving the channel close to $\vert \bm{0}_{ch}\rangle$ at all times, by either decreasing $J_0/J$~\cite{wojcik2005unmodulated,venuti2006long,venuti2007long,paganelli2013routing,bayat2015measurement} or increasing $B_0/J$~\cite{lorenzo2013quantum,apollaro2015many}. The second step is to separate the communication between each of the $M$ pairs, by tuning $B_\alpha$'s individually. In the following we, first, restrict ourselves to the case of two pairs, i.e. $M{=}2$, and consider three different strategies to maximize $\overline{F}_{\alpha\alpha}$ with minimum crosstalk. Then, we extend the results to larger $M$.


\subsection{Strategy 1 ($B_0{=}0$)}
In the first scenario, inspired by Ref.~\cite{wojcik2005unmodulated,venuti2006long,venuti2007long,paganelli2013routing,bayat2015measurement} for single user end-to-end communication, we put $B_{0}{=}0$ and consider $J_{0}{\ll} J$.
This choice of parameters creates an effective direct interaction between the sender subspace $\{S_1,S_2\}$ and the receiver ones $\{R_1,R_2\}$. To suppress the crosstalk and block the flow of information between the two subspaces of $\{S_1,R_1\}$ and $\{S_2,R_2\}$ we apply external fields $B_1$ and $B_2$ to make them energetically off-resonant from each other. By sitting at the site $R_1$ one can see the information arrives from both senders. 
In Fig.~\ref{fig:S1}(a) we plot the average transmission fidelity $\mathcal{F}_T{=}(\overline{F}_{11}+\overline{F}_{22})/2$ as function of time in a chain of $N=20$ when the parameters are tuned to $B_1/J{=}0.35$, $B_2/J{=}-0.25$ and $J_0/J{=}0.04$. As the figure shows, the average transmission fidelity evolves and at a certain time $t=\tau$, it peaks to a very high value.
In practice, at $t=\tau$ the receivers need to decouple their qubits from the data-bus or equivalently swap the quantum state from the receiver site to their registers. However, if this decoupling procedure or performing the swap operator happens at a slightly different time then the fidelity may not be at its maximum. However, this error can largely be corrected. The fast oscillations in the average transmission fidelity are due to local magnetic field $B_{\alpha}$'s and the slow dynamics following the envelope of the curve is due to the main Hamiltonian. If the decoupling procedure has a small time delay of $\Delta t$ then the error is mainly due to fast oscillation and a local rotation of the form $e^{iB_{\alpha}\sigma^{z}\Delta t}$ on site $R_{\alpha}$ largely compensates this time delay as it cancels the effect of local rotation by $B_{\alpha}$.
Interestingly, as Fig.~\ref{fig:S1}(a) shows, the crosstalks $\mathcal{F}_C{=}(\overline{F}_{12}+\overline{F}_{21})/2$ remain low and oscillate around $0.5$ resulting in negligible crosstalk between the two communicating parties.
Apart from the average fidelity one may also consider the best/worst cases among all possible states which is discussed in details in Appendix B.
In order to optimize the parameters, one can fix a time window, e.g. we choose $[1,500]{/}J$, for the dynamics of the system and then 
find optimal values for all the Hamiltonian parameters (namely $J_0^{opt}$, $B_1^{opt}$ and $B_2^{opt}$)  as well as the time $\tau$ at which all receivers should take their quantum states simultaneously. The corresponding transmission fidelities, for optimal parameters, are $\overline{F}_{\alpha\alpha}^{max}{=}\overline{F}_{\alpha\alpha}(\tau)$. 
In Fig.~\ref{fig:S1}(b) we plot $\mathcal{F}_{T}^{max}{=}(\overline{F}_{11}^{max}+\overline{F}_{22}^{max})/2$ as well as $\mathcal{F}_{C}(\tau){=}(\overline{F}_{21}(\tau)+\overline{F}_{12}(\tau))/2$ as  functions of $N$. Remarkably, for all channels of length $N{<}40$ the fidelity $\mathcal{F}_T^{max}$ remains above $0.95$, while $\mathcal{F}_{C}(\tau)$ remains around $0.5$ showing negligible crosstalks.   
At the chosen time window, the optimal coupling is obtained as $J_0^{opt}/J{=}0.04$ for all values of $N$ and its weakly depending on the length is consistent with the results of Ref.~\cite{bayat2015measurement}.
For the sake of completeness, the optimal time $\tau$ and the optimal local fields $B_1^{opt}{/}J{\in}[0.05,0.5]$, $B_2^{opt}{/}J{\in}[{-}0.5,{-}0.05]$ for any given system size $N$ are reported in Figs.~\ref{fig:S1}(c) and (d), respectively.
The local magnetic fields are chosen from intervals with opposite signs to maximize their difference while keeping their amplitude small. The optimal values for the local fields are not monotonic for different system sizes making the behavior of $\tau$ slightly irregular too.

\begin{figure}[t!]
\includegraphics[width=0.999\linewidth]{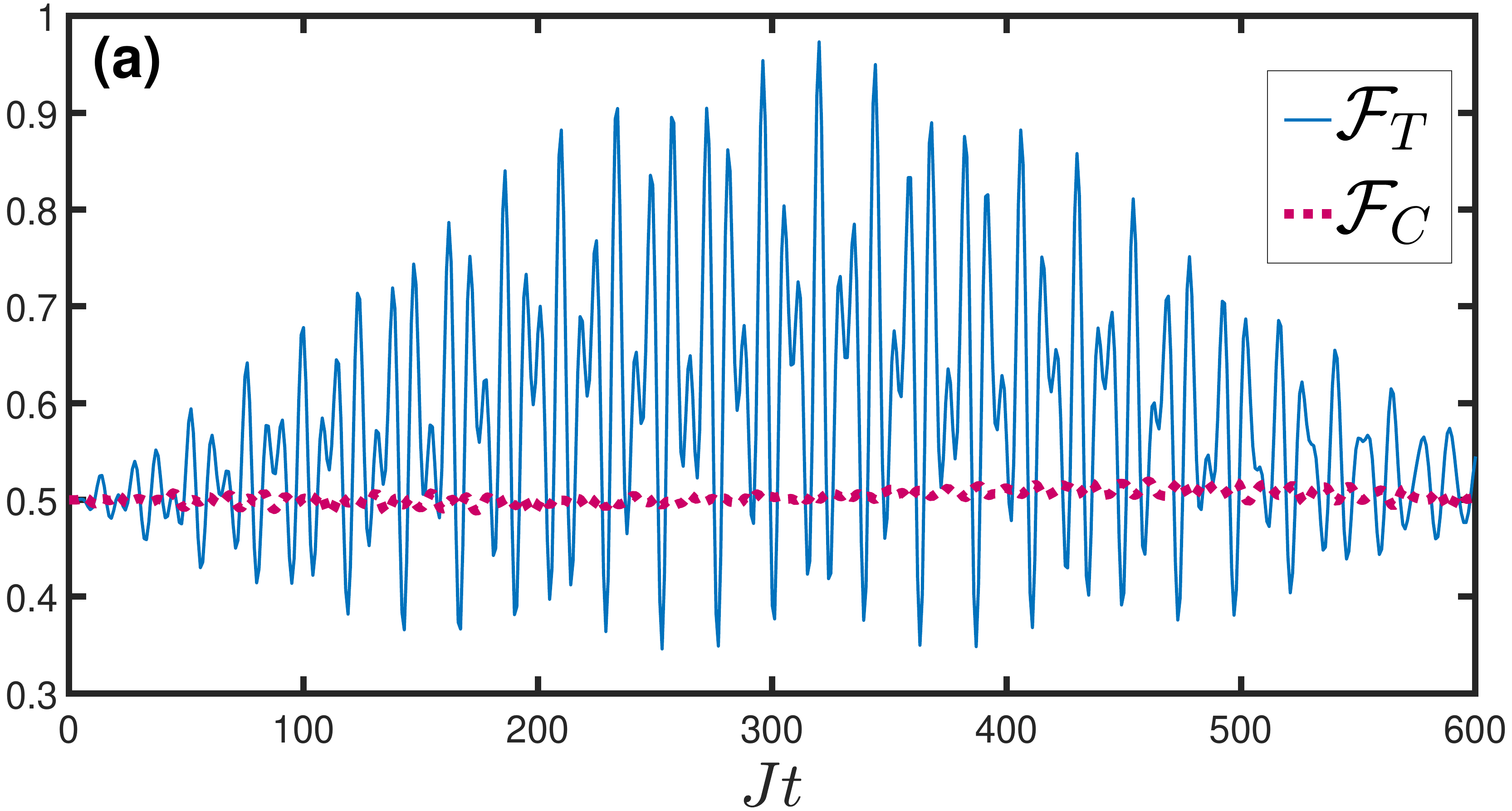}
\includegraphics[width=\linewidth , height=2.7cm]{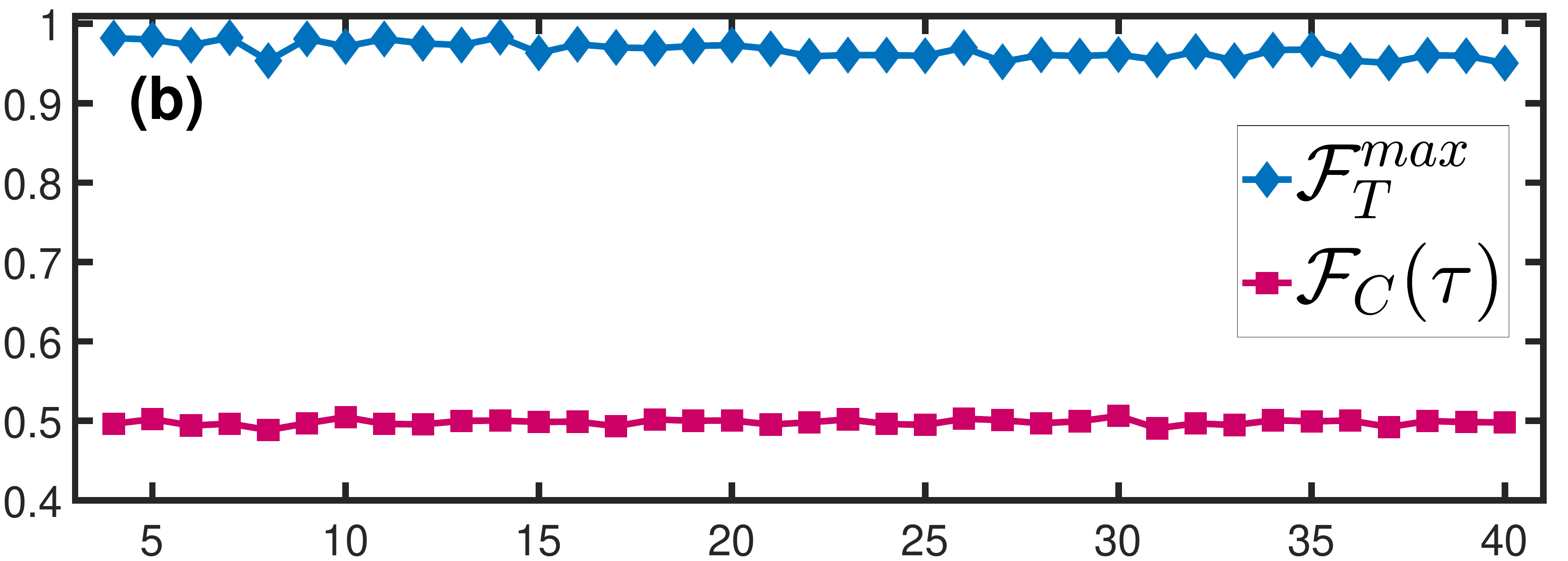}
\includegraphics[width=\linewidth , height=2.9cm]{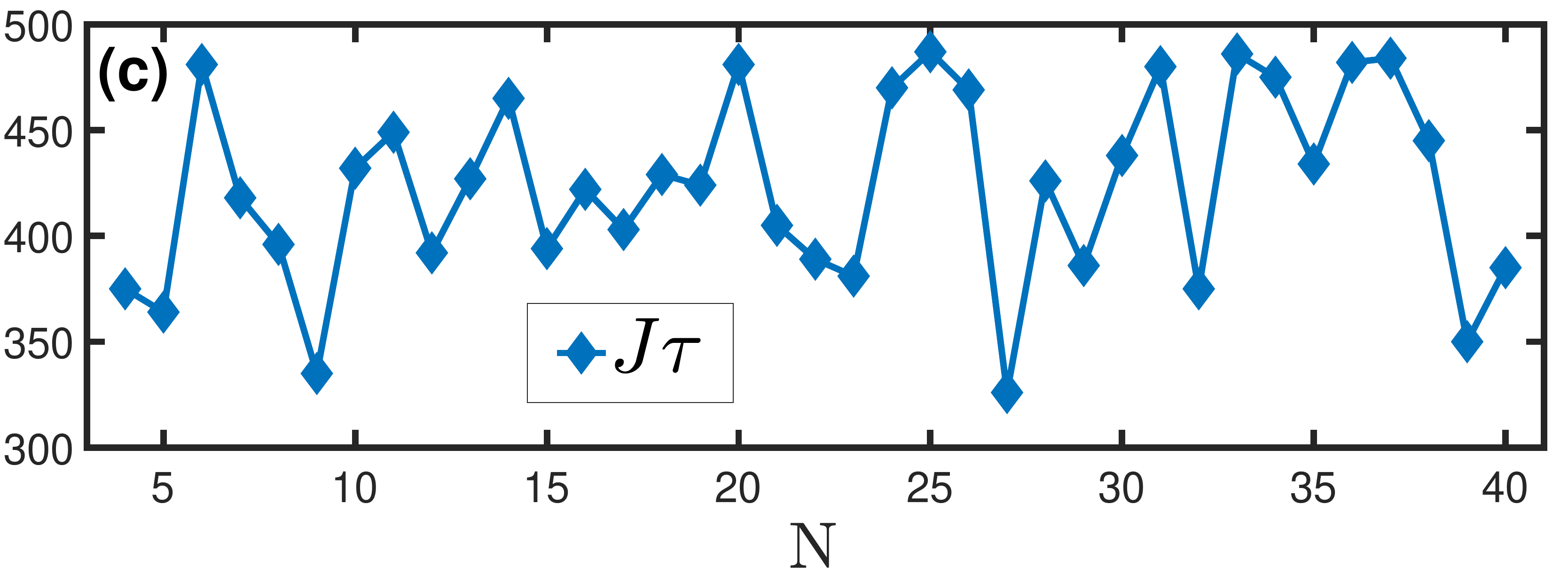}
\includegraphics[width=\linewidth , height=2.7cm]{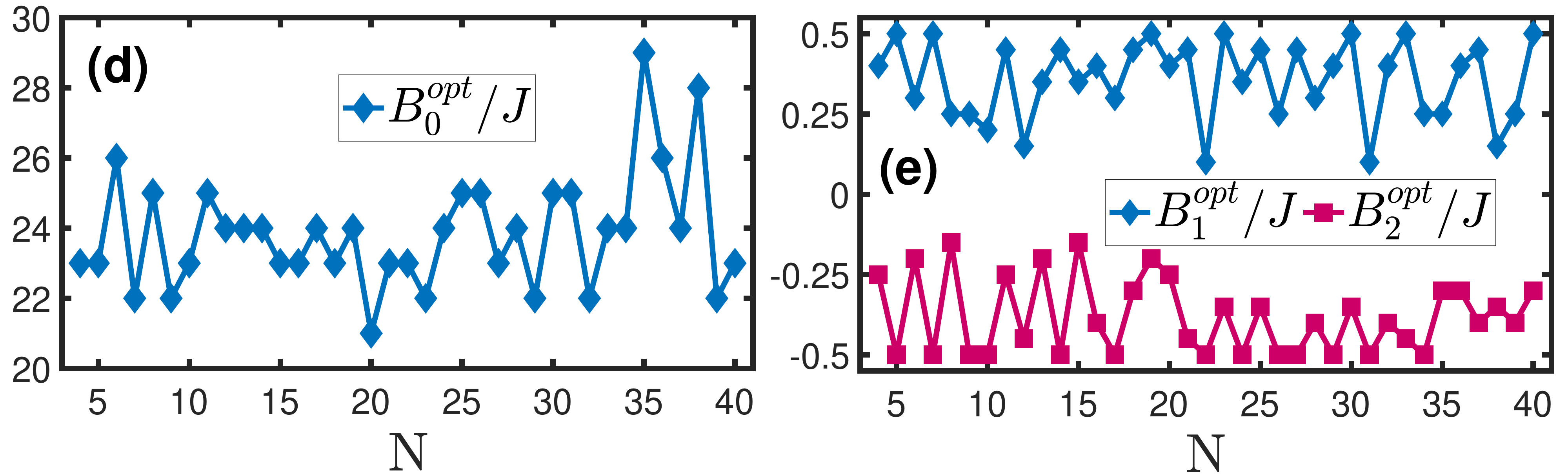}
\caption{Strategy 2: (a) The average transmission fidelity $\mathcal{F}_{T}$ and $\mathcal{F}_{C}$ as functions of time in a chain of $N{=}20$. The Hamiltonian parameters are taken as $B_{0}{/}J{=}21$, $B_{1}/J{=}0.3$, and $B_{2}/J{=}-0.35$.
(b) The obtainable average transmission fidelity $\mathcal{F}_{T}^{max}$ and  corresponding crosstalk $\mathcal{F}_{C}(\tau)$ as functions of $N$. (c) The optimal time $\tau$, optimized over the interval $\tau{\in}[1,500]/J$, at which the average transmission fidelity peaks. (d) The optimal parameter $B_{0}^{opt}/J{\in}[1,40]$. (e) The optimal parameters $B_{1}^{opt}/J{\in}[0.05,0.5]$ and $B_{2}^{opt}/J{\in}[-0.5,-0.05]$. All the plotted quantities are dimensionless.}\label{fig:S2}
\end{figure}

\subsection{Strategy 2 ($J_0{=}1$)}
Our second strategy is adopted from~\cite{lorenzo2013quantum,apollaro2015many} and is accomplished by applying a strong field $B_0$ on the ending sites of the channel and instead keep the couplings uniform, i.e.~$J_0{=}J$~(see Fig.~\ref{fig:Configuration}). 
To see the attainable fidelities for this strategy we plot  $\mathcal{F}_T$ and  $\mathcal{F}_C$ as functions of time in Fig.~\ref{fig:S2}(a) for a chain of $N{=}20$ in which $B_0/J{=}21$, $B_{1}/J{=}0.3$, and $B_{2}/J{=}-0.35$. As the figure shows  $\mathcal{F}_T$ reaches very high values and peaks at $t{=}\tau$. Remarkably,  $\mathcal{F}_C$ fluctuates around $0.5$, showing very small crosstalks.
Analogous to the previous strategy, one can optimize the time of the evolution as well as the Hamiltonian parameters within a chosen time window, here again $[0,500/J]$. In Fig.~\ref{fig:S2}(b) we report the maximum of the average transmission fidelity $\mathcal{F}_T^{max}$ and the average crosstalk $\mathcal{F}_C(\tau)$ as functions of $N$. This figure shows that while the transmission fidelities for both parties achieve above $0.95$ the crosstalks between them remain negligible. The optimal time $\tau$ for obtaining such quantities is plotted in Fig.~\ref{fig:S2}(c).
The reason that the optimal times oscillate with length is because the chosen time window allows for several peaks and their maximum changes as the length vary. 
The other optimal parameters such as $B_0^{opt}{/}J{\in}[20,40]$ and as well as $B_1^{opt}{/}J{\in}[0.05,0.5]$, $B_2^{opt}{/}J{\in}[-0.5,-0.05]$ are presented in Figs.~\ref{fig:S2}(d) and (e), respectively. The results show that by tuning $20 {\le} B_0/J {\le} 30$ one needs weak local magnetic fields $B_\alpha$'s for obtaining high fidelity simultaneous transmission.


\subsection{Strategy 3}    
The third scenario is a hybrid of both outlined strategies and the performance of the channel is investigated when both $B_0$ and $J_0$ are optimized. Again we fix the time window to $[1,500]{/}J$ and optimize the time and the parameters $B_0$, $J_0$, $B_1$, and $B_2$ to maximize the average transmission fidelity and keeping the average crosstalk negligible.    
In TABLE~\ref{table1} we report the maximum fidelity $\mathcal{F}_T^{max}$, the corresponding crosstalk $\mathcal{F}_C(\tau)$, the optimal time $\tau$ as well as the optimized values of the Hamiltoninan parameters for different values of $N$.
Clearly, $J_0^{opt}$ and $B_0^{opt}$ are midway between the two previous strategies, namely $J_0^{opt}$ becomes larger in comparison with the optimal values in strategy $1$ and $B_0^{opt}$ becomes smaller than the case of strategy $2$.
A comparison between different strategies shows that, for long chains strategy $3$ is superior to the other ones in terms of fidelity, indicating that a hybrid optimization of both $B_0$ and $J_0$ outperforms the optimization of individual parameters.
\begin{table}[h!]
{
\renewcommand{\arraystretch}{1.35}
\setlength{\tabcolsep}{3pt}
\begin{tabular}{| m{0.6cm} |  m{1cm}  m{1cm}  m{1cm}  m{1cm}   m{1cm}  m{1cm} m{1cm} m{0.9cm} |} 
\hline 
\multicolumn{1}{|c|}{N}& \multicolumn{1}{c}{5} & \multicolumn{1}{c}{10} & \multicolumn{1}{c}{15} & \multicolumn{1}{c}{20} & \multicolumn{1}{c}{25} & \multicolumn{1}{c}{30} & \multicolumn{1}{c}{35} & \multicolumn{1}{c|}{40}
\\  \hline \hline
\multicolumn{1}{|c|}{$\mathcal{F}_T^{max}$} &\multicolumn{1}{c}{$0.990$} & \multicolumn{1}{c}{$0.983$} & \multicolumn{1}{c}{$0.977$} & \multicolumn{1}{c}{$0.977$} & \multicolumn{1}{c}{$0.971$}&\multicolumn{1}{c}{$0.966$}& \multicolumn{1}{c}{$0.975$} & \multicolumn{1}{c|}{$0.965$}
\\  \hline
\multicolumn{1}{|c|}{$\mathcal{F}_C(\tau)$} &\multicolumn{1}{c}{$0.498$} & \multicolumn{1}{c}{$0.499$} & \multicolumn{1}{c}{$0.501$} & \multicolumn{1}{c}{$0.496$} & \multicolumn{1}{c}{$0.497$}&\multicolumn{1}{c}{$0.506$}& \multicolumn{1}{c}{$0.501$} & \multicolumn{1}{c|}{$0.499$}
\\  \hline
\multicolumn{1}{|c|}{$\tau$} &\multicolumn{1}{c}{$463$} & \multicolumn{1}{c}{$419$} & \multicolumn{1}{c}{$474$} & \multicolumn{1}{c}{$488$} & \multicolumn{1}{c}{$492$}&\multicolumn{1}{c}{$499$}& \multicolumn{1}{c}{$469$} & \multicolumn{1}{c|}{$500$}
\\  \hline \hline
\multicolumn{1}{|c|}{$J_0^{opt}/J$} &\multicolumn{1}{c}{$0.04$} & \multicolumn{1}{c}{$0.04$} & \multicolumn{1}{c}{$0.8$} & \multicolumn{1}{c}{$0.8$} & \multicolumn{1}{c}{$0.8$}&\multicolumn{1}{c}{$0.85$}& \multicolumn{1}{c}{$0.85$} & \multicolumn{1}{c|}{$0.7$}
\\  \hline
\multicolumn{1}{|c|}{$B_0^{opt}/J$} &\multicolumn{1}{c}{$0$} & \multicolumn{1}{c}{$0$} & \multicolumn{1}{c}{$20$} & \multicolumn{1}{c}{$20$} & \multicolumn{1}{c}{$21$}&\multicolumn{1}{c}{$25$}& \multicolumn{1}{c}{$24$} & \multicolumn{1}{c|}{$21$}
\\  \hline
\multicolumn{1}{|c|}{$B_1^{opt}/J$} &\multicolumn{1}{c}{$0.35$} & \multicolumn{1}{c}{$0.45$} & \multicolumn{1}{c}{$-0.25$} & \multicolumn{1}{c}{$-0.3$} & \multicolumn{1}{c}{$-0.35$}&\multicolumn{1}{c}{$-0.25$}& \multicolumn{1}{c}{$-0.3$} & \multicolumn{1}{c|}{$-0.35$}
\\  \hline
\multicolumn{1}{|c|}{$B_2^{opt}/J$} &\multicolumn{1}{c}{$-0.5$} & \multicolumn{1}{c}{$-0.1$} & \multicolumn{1}{c}{$0.2$} & \multicolumn{1}{c}{$0.4$} & \multicolumn{1}{c}{$0.45$}&\multicolumn{1}{c}{$0.4$}& \multicolumn{1}{c}{$0.25$} & \multicolumn{1}{c|}{$0.45$}
\\  \hline
\end{tabular}
\caption{Strategy 3: The maximum of $\mathcal{F}_T$ and corresponding $\mathcal{F}_C(\tau)$ in optimal time $\tau$ for strategy $3$ in chains with different lengths. The optimal exchange coupling, $J_0^{opt}/J{\in}[0.01,1]$, the optimal local magnetic field on the ends of the chain, $B_0^{opt}/J{\in}[1,40]$ and the optimal values of the local fields on users' qubits, $B_1^{opt}/J{\in}[-0.5,0.5]$, $B_2^{opt}/J{\in}[-0.5,0.5]$ for providing the presented results.}\label{table1}
}
\end{table}


\section{Multiple users}
The proposed protocol with all the three strategies can be generalized to more than two users.  
No matter how many users we consider, one can always tune the parameters to keep the crosstalks negligible. 
To confirm this expectation, we study the performance of two strategies $1$ and $2$ in the case of three users.
Our results show that in different spin chains, three users can simultaneously communicate with the average transmission fidelity $\mathcal{F}_T{=}\sum_{\alpha=1}^{3}\overline{F}_{\alpha\alpha}/3$ more than $0.94$ while keeping the average crosstalk  $\mathcal{F}_C{=}\sum_{\alpha\neq\beta=1}^{3}\overline{F}_{\alpha\beta}/6$ around $0.5$ within the time scale of $[1,500]{/}J$. 
In TABLE~\ref{table} we report $\mathcal{F}^{max}_T$ and $\mathcal{F}_C(\tau)$, the optimal time $\tau$ and also corresponding optimal parameters for some system sizes by adopting the first and second strategies. The transmission fidelities remain steadily high and comparable with the case of two users. Interestingly the optimal coupling strength in the first strategy is obtained as $J_{0}^{opt}/J{=}0.04$ for all considered chains which is very close to the case of $2$ users.

\begin{table*}[t]
{
\renewcommand{\arraystretch}{1.35}
\setlength{\tabcolsep}{2.5pt}
\begin{tabular}{| m{2mm}|  m{1cm} |  m{1cm}  m{1cm}  m{1cm}  m{1cm}   m{1cm}  m{1cm}  m{0.9cm} |} 
\hline 
\multicolumn{2}{|c|}{N}& \multicolumn{1}{c}{5} & \multicolumn{1}{c}{6} & \multicolumn{1}{c}{7} & \multicolumn{1}{c}{8} & \multicolumn{1}{c}{9} & \multicolumn{1}{c}{10} & \multicolumn{1}{c|}{20} 
\\  \hline \hline
\multicolumn{1}{|c|}{\multirow{7}{*}{\rotatebox[origin=c]{90}{{$Strategy$ $1$}}}} & \multicolumn{1}{c|}{$\mathcal{F}_T^{max}$} & \multicolumn{1}{c}{$0.975$} & \multicolumn{1}{c}{$0.961$} & \multicolumn{1}{c}{$0.984$} & \multicolumn{1}{c}{$0.960$}&\multicolumn{1}{c}{$0.966$}& \multicolumn{1}{c}{$0.966$} &\multicolumn{1}{c|}{$0.958$}
\\  \cline{2-9} 
\multicolumn{1}{|c|}{}& \multicolumn{1}{c|}{$\mathcal{F}_C(\tau)$} & \multicolumn{1}{c}{$0.499$} & \multicolumn{1}{c}{$0.497$} & \multicolumn{1}{c}{$0.499$} & \multicolumn{1}{c}{$0.500$}&\multicolumn{1}{c}{$0.496$}& \multicolumn{1}{c}{$0.501$} &                    \multicolumn{1}{c|}{$0.498$}
\\  \cline{2-9} 
\multicolumn{1}{|c|}{}& \multicolumn{1}{c|}{$\tau$ } & \multicolumn{1}{c}{$446$} & \multicolumn{1}{c}{$438$} & \multicolumn{1}{c}{$474$} & \multicolumn{1}{c}{$428$}&\multicolumn{1}{c}{$447$}& \multicolumn{1}{c}{$438$} &
\multicolumn{1}{c|}{$435$}
\\  \cline{2-9}
\multicolumn{1}{|c|}{\multirow{4}{*}{}} & \multicolumn{1}{c|}{$J_{0}^{opt}/J$}  & \multicolumn{1}{c}{$0.04$} & \multicolumn{1}{c}{$0.04$} & \multicolumn{1}{c}{$0.04$} & \multicolumn{1}{c}{$0.04$}&\multicolumn{1}{c}{$0.04$}& \multicolumn{1}{c}{$0.04$} &\multicolumn{1}{c|}{$0.04$}
\\  \cline{2-9}
\multicolumn{1}{|c|}{\multirow{3}{*}{}} & \multicolumn{1}{c|}{$B_{1}^{opt}/J$}  & \multicolumn{1}{c}{$-0.4$} & \multicolumn{1}{c}{$0.1$} & \multicolumn{1}{c}{$-0.5$} & \multicolumn{1}{c}{$-0.5$}&\multicolumn{1}{c}{$-0.45$}& \multicolumn{1}{c}{$-0.5$} &\multicolumn{1}{c|}{$-0.35$}
\\  \cline{2-9} 
\multicolumn{1}{|c|}{}& \multicolumn{1}{c|}{$B_{2}^{opt}/J$} & \multicolumn{1}{c}{$-0.3$} & \multicolumn{1}{c}{$0.15$} & \multicolumn{1}{c}{$-0.4$} & \multicolumn{1}{c}{$0.15$}&\multicolumn{1}{c}{$-0.35$}& \multicolumn{1}{c}{$0.1$} &                    \multicolumn{1}{c|}{$0.25$}
\\  \cline{2-9} 
\multicolumn{1}{|c|}{}& \multicolumn{1}{c|}{$B_{3}^{opt}/J$ } & \multicolumn{1}{c}{$0.35$} & \multicolumn{1}{c}{$0.2$} & \multicolumn{1}{c}{$0.3$} & \multicolumn{1}{c}{$-0.05$}&\multicolumn{1}{c}{$-0.4$}& \multicolumn{1}{c}{$-0.45$} &
\multicolumn{1}{c|}{$-0.05$}                       
\\  \hline 
\end{tabular}
\qquad
\begin{tabular}{| m{2mm}|  m{1cm} |  m{1cm}  m{1cm}  m{1cm}  m{1cm}   m{1cm}  m{1cm}  m{0.9cm} |} 
\hline 
\multicolumn{2}{|c|}{N}& \multicolumn{1}{c}{5} & \multicolumn{1}{c}{6} & \multicolumn{1}{c}{7} & \multicolumn{1}{c}{8} & \multicolumn{1}{c}{9} & \multicolumn{1}{c}{10} & \multicolumn{1}{c|}{20} 
\\  \hline \hline
\multirow{7}{*}{\rotatebox[origin=c]{90}{{ $Strategy$ $2$}}}&
\multicolumn{1}{c|}{$\mathcal{F}_T^{max}$} & \multicolumn{1}{c}{$0.972$} & \multicolumn{1}{c}{$0.974$} & \multicolumn{1}{c}{$0.971$} & \multicolumn{1}{c}{$0.975$}&\multicolumn{1}{c}{0.967}& \multicolumn{1}{c}{$0.967$} &\multicolumn{1}{c|}{$0.949$}
\\  \cline{2-9} 
\multicolumn{1}{|c|}{}& \multicolumn{1}{c|}{$\mathcal{F}_C(\tau)$} & \multicolumn{1}{c}{$0.499$} & \multicolumn{1}{c}{$0.498$} & \multicolumn{1}{c}{$0.497$} & \multicolumn{1}{c}{$0.499$}&\multicolumn{1}{c}{$0.499$}& \multicolumn{1}{c}{$0.498$} &                    \multicolumn{1}{c|}{$0.501$}
\\  \cline{2-9} 
\multicolumn{1}{|c|}{}& \multicolumn{1}{c|}{$\tau$ } & \multicolumn{1}{c}{$459$} & \multicolumn{1}{c}{$378$} & \multicolumn{1}{c}{$491$} & \multicolumn{1}{c}{$500$}&\multicolumn{1}{c}{450}& \multicolumn{1}{c}{472} &
\multicolumn{1}{c|}{$500$}
\\  \cline{2-9}
\multicolumn{1}{|c|}{} & \multicolumn{1}{c|}{$B_{0}^{opt}/J$}  & \multicolumn{1}{c}{$26$} & \multicolumn{1}{c}{$21$} & \multicolumn{1}{c}{$27$} & \multicolumn{1}{c}{$26$}&\multicolumn{1}{c}{$25$}& \multicolumn{1}{c}{$25$} &\multicolumn{1}{c|}{$28$}
\\  \cline{2-9} 
\multicolumn{1}{|c|}{}& \multicolumn{1}{c|}{$B_{1}^{opt}/J$} & \multicolumn{1}{c}{$-1.1$} & \multicolumn{1}{c}{$-1.1$} & \multicolumn{1}{c}{$-0.6$} & \multicolumn{1}{c}{$-0.8$}&\multicolumn{1}{c}{$-0.4$}& \multicolumn{1}{c}{$-0.7$} &                    \multicolumn{1}{c|}{$-1$}
\\  \cline{2-9} 
\multicolumn{1}{|c|}{}& \multicolumn{1}{c|}{$B_{2}^{opt}/J$ } & \multicolumn{1}{c}{$0.5$} & \multicolumn{1}{c}{$0.1$} & \multicolumn{1}{c}{$0.4$} & \multicolumn{1}{c}{$0.1$}&\multicolumn{1}{c}{$1.0$}& \multicolumn{1}{c}{$0.0$} &
\multicolumn{1}{c|}{$0.6$} 
\\  \cline{2-9} 
\multicolumn{1}{|c|}{}& \multicolumn{1}{c|}{$B_{3}^{opt}/J$ } & \multicolumn{1}{c}{$1.1$} & \multicolumn{1}{c}{$1.4$} & \multicolumn{1}{c}{$1.2$} & \multicolumn{1}{c}{$1.0$}&\multicolumn{1}{c}{$0.3$}& \multicolumn{1}{c}{$1.2$} &
\multicolumn{1}{c|}{$1.2$}
\\  \hline
\end{tabular}
\caption{Three users: The maximum of $\mathcal{F}_T$ and corresponding $\mathcal{F}_C(\tau)$ in optimal time $\tau$ using strategy $1$ and $2$ in different chains. Here, the optimal exchange coupling $J_0^{opt}/J$ for strategy $1$ has been optimized over the interval $J_0^{opt}/J{\in}[0.01,1]$ and the optimal local magnetic field on the ends of the chain $B_0^{opt}/J$ for strategy $2$ has been optimized over $B_0^{opt}/J{\in}[1,40]$. In both strategies, the optimal values of the local fields on users' qubits, $B_1^{opt}/J$, $B_2^{opt}/J$ and $B_3^{opt}/J$, have been optimized over the interval $[-1.5,1.5]$}\label{table}
}
\end{table*}


\section{Bi-localized eigenstates}
The main reason behind the achievement of high transmission fidelities and low crosstalk is the emergence of bi-localized eigenstates whose excitations are mainly localized at sender and receiver sites. Since these bi-localized eigenstates are the only ones involving in the dynamics of the system, the channel mostly remains in the state $\vert\bm{0}_{ch}\rangle$. Consequently, effective end-to-end interaction is generated between the sender and receiver qubits. The emergence of bi-localized qubits is mainly due to the engineering of $J_0$ and $B_0$ and then, to minimize the crosstalk, further localizing the excitations between each pair $(S_\alpha,R_\alpha)$ is achieved by tuning $B_\alpha$'s. See Appendix B  for details.


\section{PERFORMANCE UNDER REALISTIC CONDITIONS}
In the previous sections we have illustrated that multiple users can accomplish high-fidelity simultaneous communication with negligible crosstalk by tuning Hamiltonian parameters. However, acquiring this result is based on four ideal assumptions, namely: (i) the chain is initially prepared in the state $\vert\bm{0}_{ch}\rangle$; (ii) the couplings are adjusted accurately to their specific values; (iii) the local magnetic fields can be tuned perfectly; and (iv) the system is isolated from its environment. In this section, we investigate imperfect scenarios in which these assumptions are relaxed. For the sake of brevity and without loss of generality, we focus on two-user communication considering only our first and second strategies. Since the third strategy is a hybrid of the first two, the impact of the imperfections will approximately be the average of the impacts on the first two strategies.

\subsection{Thermal Initial State}

In practice, thermal fluctuations may create excitation in the channel. To investigate the effect of finite temperatures, we consider the initial state of the channel to take the form of a thermal ensemble
\begin{equation}\label{eq:Thermal State}
\rho_{ch}=\dfrac{e^{-H_{ch}/K_{B}T}}{Tr[e^{-H_{ch}/K_{B}T}]},
\end{equation}
where $T$ is temperature and $K_{B}$ is the Boltzmann constant. 
To see the impact of finite temperature $T$, we compute the average fidelity $\overline{F}_{\alpha\beta}(t)$, for which we provide a compact form in Appendix A.
Fig.~\ref{fig:thermalState} shows the maximum of transmission fidelities in a chain of length $N=6$ for our first two strategies.
As the figure shows, by increasing the temperature the fidelity first remains very high, showing a plateau at small temperatures, and then monotonically decreases to eventually reach the classical threshold of $2/3$ for transferring quantum information~\cite{bose2003quantum}. The width of the plateau is determined by the energy gap of the finite system and is consistent with previous observations~\cite{bayat2010information}.  Interestingly, Fig.~\ref{fig:thermalState} shows that while the strategy $1$ gives higher fidelity at low temperatures, in higher temperatures it is the strategy $2$ that gives better transmission quality. Therefore, depending on the temperature of the system one strategy may result in a higher fidelity than the other.

\begin{figure}[t!]
\centering\offinterlineskip
\includegraphics[width=\linewidth]{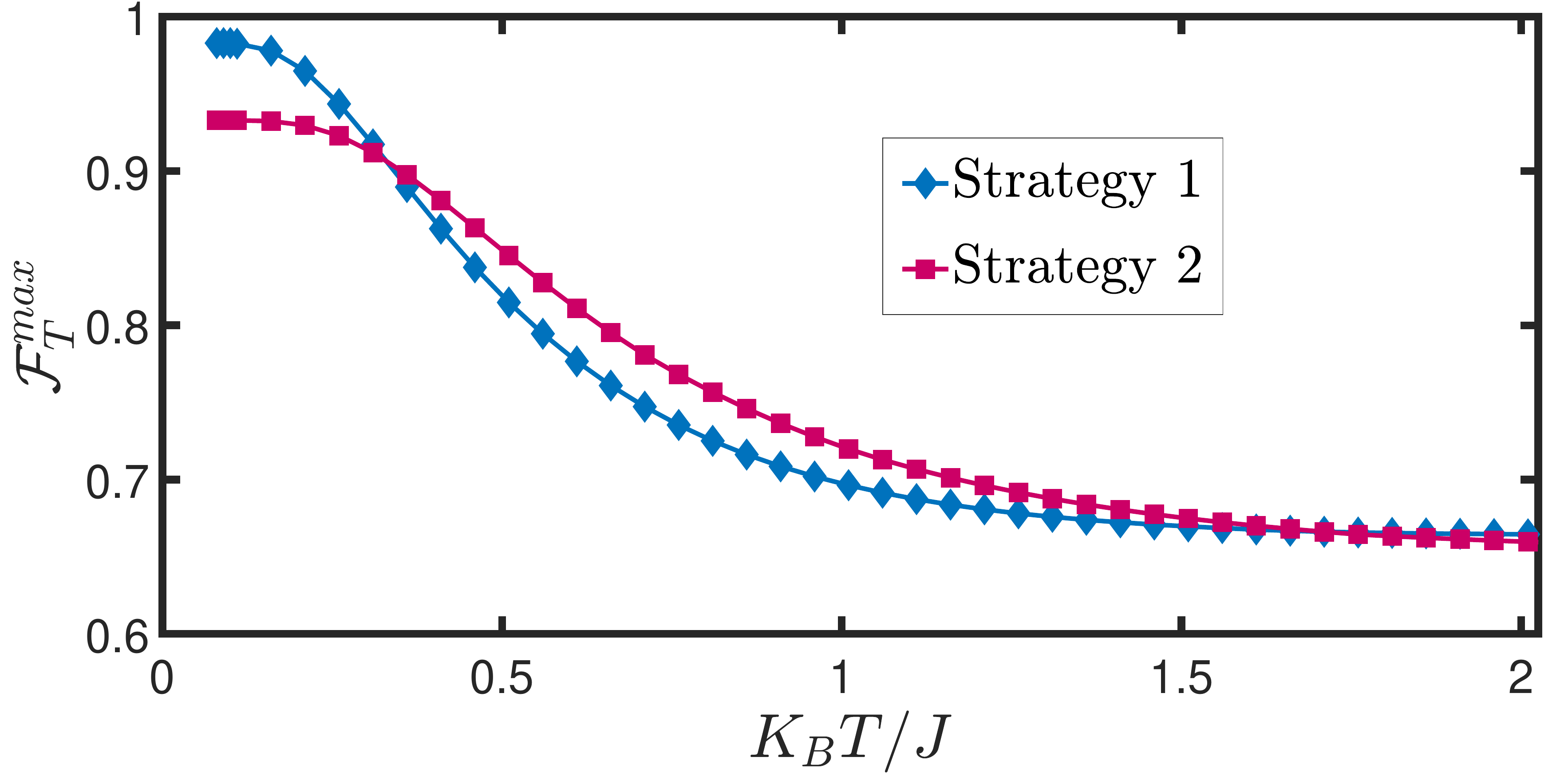} 
\caption{Thermal Initial State: The maximum of transmission fidelities for strategies $1$ and $2$ as function of dimensionless parameter $K_{B}T/J$ in a spin chain of length $N=6$. In preparing these plots the Hamiltonian parameters are tuned as $\{J_0/J=0.04, B_1/J=0.15, B_2/J=-0.05\}$ and $\{B_0/J=26, B_1/J=0.3, B_2/J=-0.2\}$, respectively, for strategies $1$ and $2$. }\label{fig:thermalState}
\end{figure}


\subsection{Random Coupling}  
 
\begin{figure}[t]
\centering\offinterlineskip
\includegraphics[width=\linewidth,height=4cm]{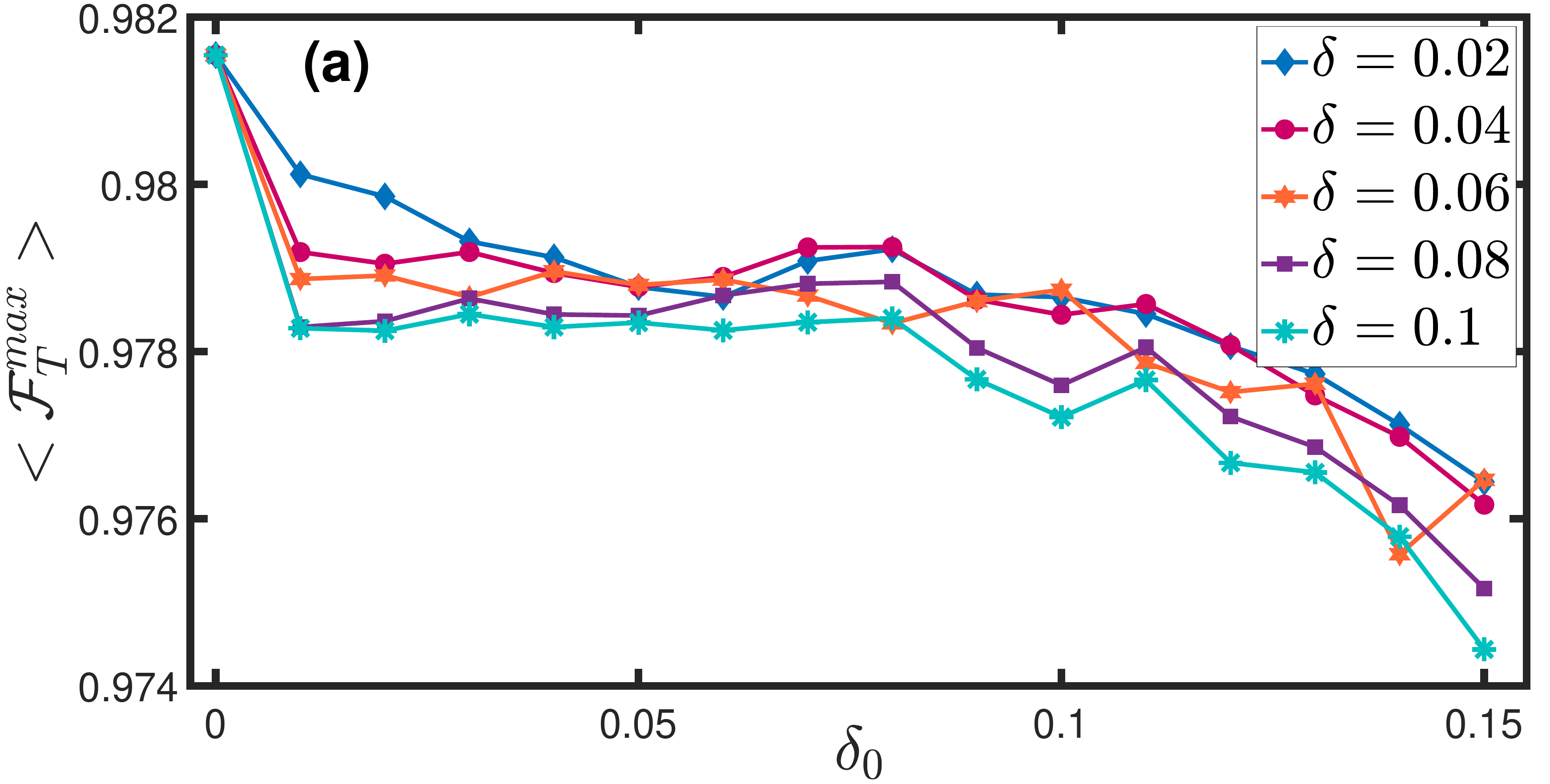} 
\includegraphics[width=\linewidth,height=4cm]{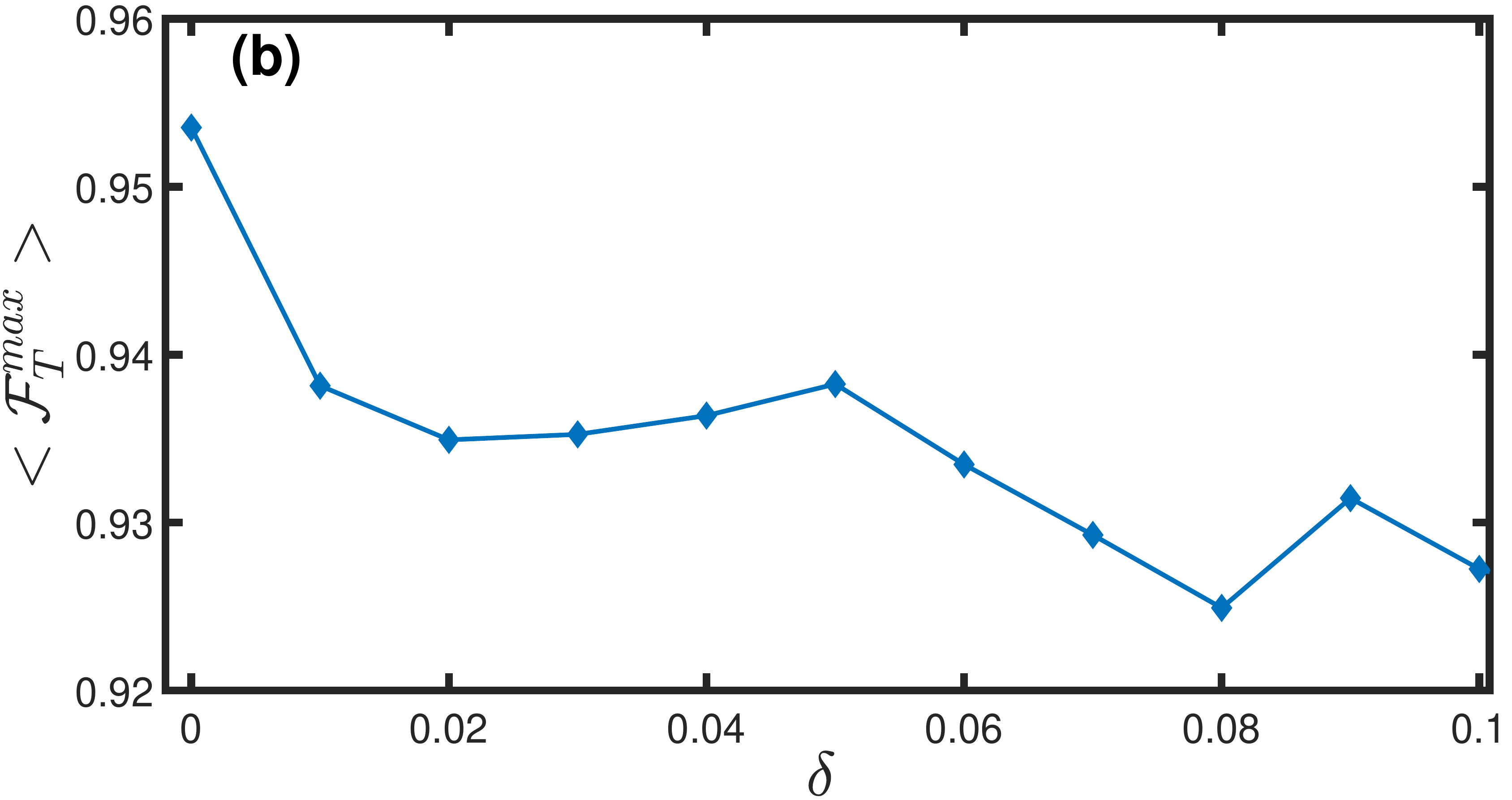}
\caption{Random Coupling: (a) The maximum average transmission fidelity for various values of $\delta$ as function of $\delta_{0}$ in a spin chain of length $N=8$ for strategy $1$. The Hamiltonian parameters are taken as $J_0/J=0.04$, $B_1/J=0.15$ and  $B_2/J=-0.05$.  (b) The maximum average transmission fidelity for strategy $2$, as function of $\delta_{0}$ in a spin chain with $N=8$. 
The Hamiltonian parameters are tuned as $B_0/J=25$, $B_1/J=0.25$ and $B_2/J=-0.15$. All the plotted quantities are dimensionless.}\label{fig:randomcoupling}
\end{figure}

The second assumption in our protocols is the homogeneity of the Hamiltonian $H_{ch}$ and tunability of $J_{0}$. However, the exchange couplings may not be as precise as we expect and random variations are inevitable during fabrication. For investigating the effect of such randomness on the quality of our protocols, we assume that the first terms of Eq.~(\ref{eq:channel Hamiltonian}) and Eq.~(\ref{eq:Interaction Hamiltonian}) are updated as $J\textstyle \sum_{i=1}^{N-1}(1+j_{i})(\sigma_{i}^{x}\sigma_{i+1}^{x}+\sigma_{i}^{y}\sigma_{i+1}^{y})$ and $J_0 \textstyle\sum_{\alpha=1}^{M}(1+j_{0\alpha})\left(\sigma_{S_\alpha}^{x}\sigma_{1}^{x}+\sigma_{S_\alpha}^{y}\sigma_{1}^{y}+\sigma_{N}^{x}\sigma_{R_\alpha}^{x}+\sigma_{N}^{y}\sigma_{R_\alpha}^{y} \right)$, respectively.
Here, $j_{i}\in[-\delta,+\delta]$ and $j_{0\alpha}\in[-\delta_{0},+\delta_{0}]$ are uniformly distributed random variables with zero means. We generate $100$ different random Hamiltonians, according to these distributions, for each values of $\delta$ and $\delta_{0}$, and obtain the maximum average fidelity $\mathcal{F}_{T}^{max}$. By averaging over all these random realizations  one gets $\langle \mathcal{F}_{T}^{max} \rangle$ as a parameter to quantify the quality of transfer. In Fig.~\ref{fig:randomcoupling}(a), we depict the results for different values of $\delta$ as function of $\delta_{0}$ in a spin chain of length $N=8$ when strategy $1$ is adopted. The protocol shows 
very robust behavior, even for a strong disorder with strength $\delta_0=0.15$.  In Fig.~\ref{fig:randomcoupling}(b) we plot the transmission fidelity $\langle \mathcal{F}_{T}^{max} \rangle$  as a function of $\delta$ in a chain of length $N=8$ when the strategy $2$ is adopted. In compare to the strategy 1, the fidelity is more susceptible to randomness and thus decays faster. Nonetheless, for even a strong disorder with strength $\delta=0.1$ the fidelity $\langle \mathcal{F}_{T}^{max} \rangle$ still remains above $0.92$. 
      
\begin{figure}[t]
\centering\offinterlineskip
\includegraphics[width=\linewidth,height=4cm]{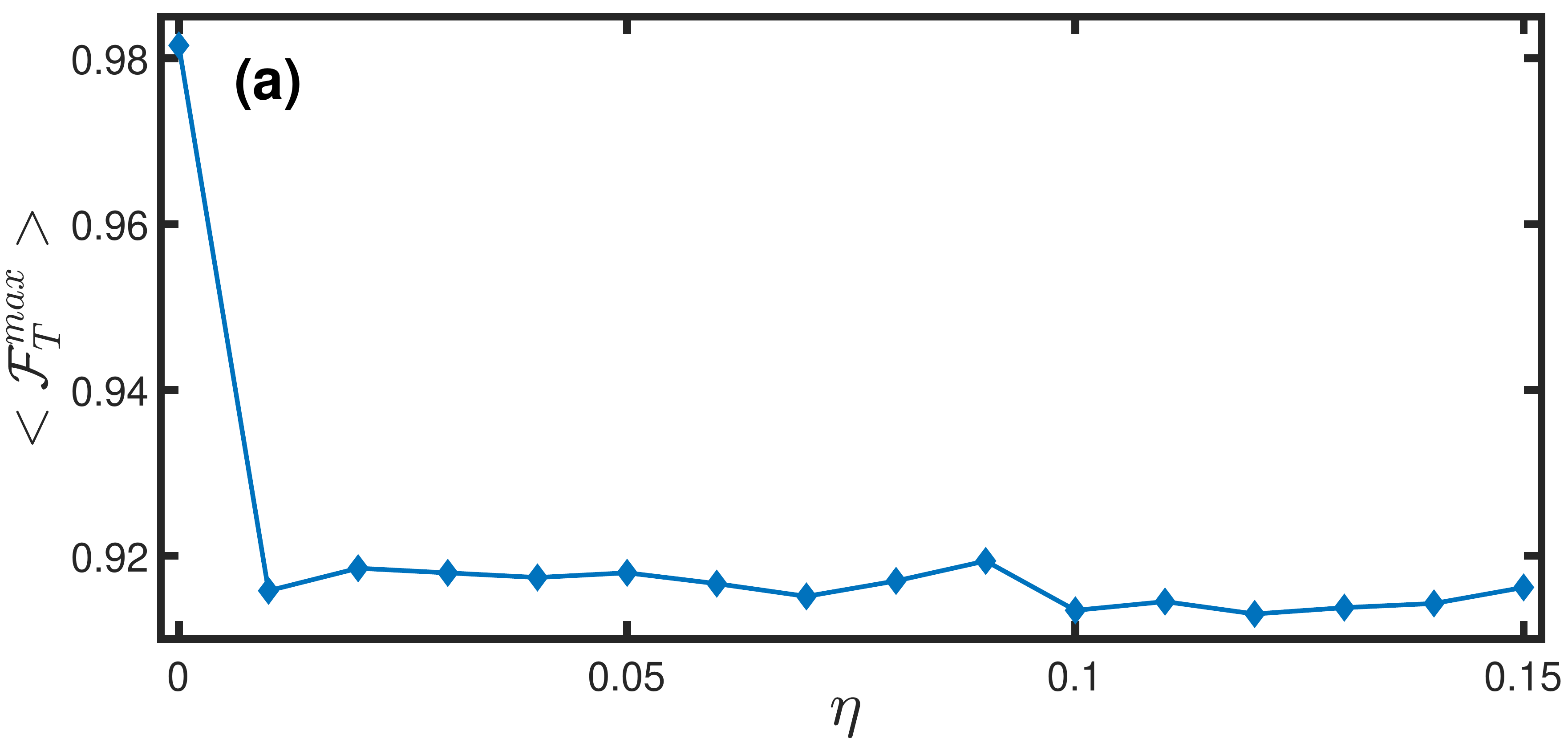} 
\includegraphics[width=\linewidth,height=4cm]{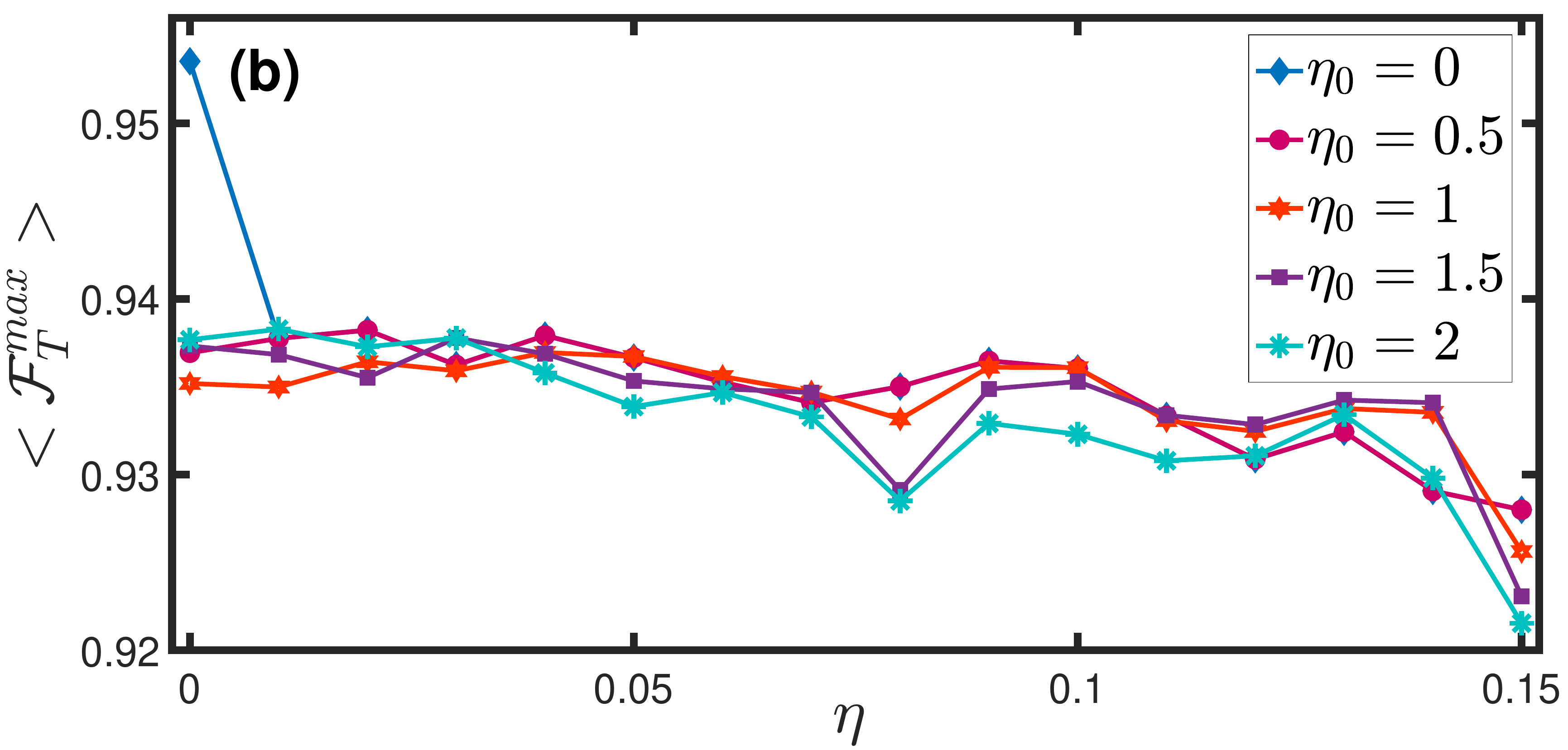}
\caption{Inaccurate local magnetic fields: (a) The maximum average transmission fidelity for strategy $1$, as function of $\eta$ in a chain with $N=8$ spins. The Hamiltonian Parameters are adjusted as $J_0/J{=}0.04$, $B_1/J{=}0.15$ and $B_2/J{=}-0.05$. (b) The maximum average transmission fidelity for various values of $\eta_{0}$ as function of $\delta_{0}$ in a spin chain with $N=8$ by adopting strategy $2$. The Hamiltonian parameters are tuned as $B_0/J=25$, $B_1/J=0.25$ and $B_2/J=-0.15$. All the plotted quantities are dimensionless.}\label{fig:magneticfield}
\end{figure}

\subsection{Inaccurate Local Magnetic Fields}
The key point for the success of our protocol is to properly adjust the local magnetic fields, namely $B_{\alpha}$'s and $B_0$. 
The inaccuracy in tuning these fields may affect the obtainable fidelities. To investigate this effect, analogous to the previous section, we assume $B_{0}$ and $B_{\alpha}$, are random variables that vary around average values $B_{0}^{opt}$ and $B_{\alpha}^{opt}$, respectively. So, the coefficient of the second terms of Eq.~(\ref{eq:channel Hamiltonian}) and Eq.~(\ref{eq:Interaction Hamiltonian}) are considered to be  $B_0^{opt}(1+b_{0})$ and $B_{\alpha}^{opt}(1+b_{\alpha})$,  respectively. Where $b_0\in [-\eta_0,\eta_0]$ and $b_{\alpha}\in [-\eta,\eta]$ are uniformly distributed random variables with zero means. For each values of $\eta_0$ and $\eta$ we repeat the procedure for $100$ random realizations to get the average fidelity $\langle \mathcal{F}_{T}^{max} \rangle$. In Fig.~\ref{fig:magneticfield}(a) the average transmission fidelity $\langle \mathcal{F}_{T}^{max} \rangle$ is plotted as a function of $\eta$ for a chain of length $N=8$ using our first strategy. The fidelity shows fairly stable behavior and remains almost steady around $0.92$ after a short decay.
In Fig.~\ref{fig:magneticfield}(b) we plot the same quantity for our second strategy for various choices of $\eta_0$. Again the fidelity has a stable behavior against disorder in magnetic field, even up to $\eta=0.15$.

\subsection{Dephasing} 
\begin{figure}[t!]
\centering\offinterlineskip
\includegraphics[width=\linewidth,height=4cm]{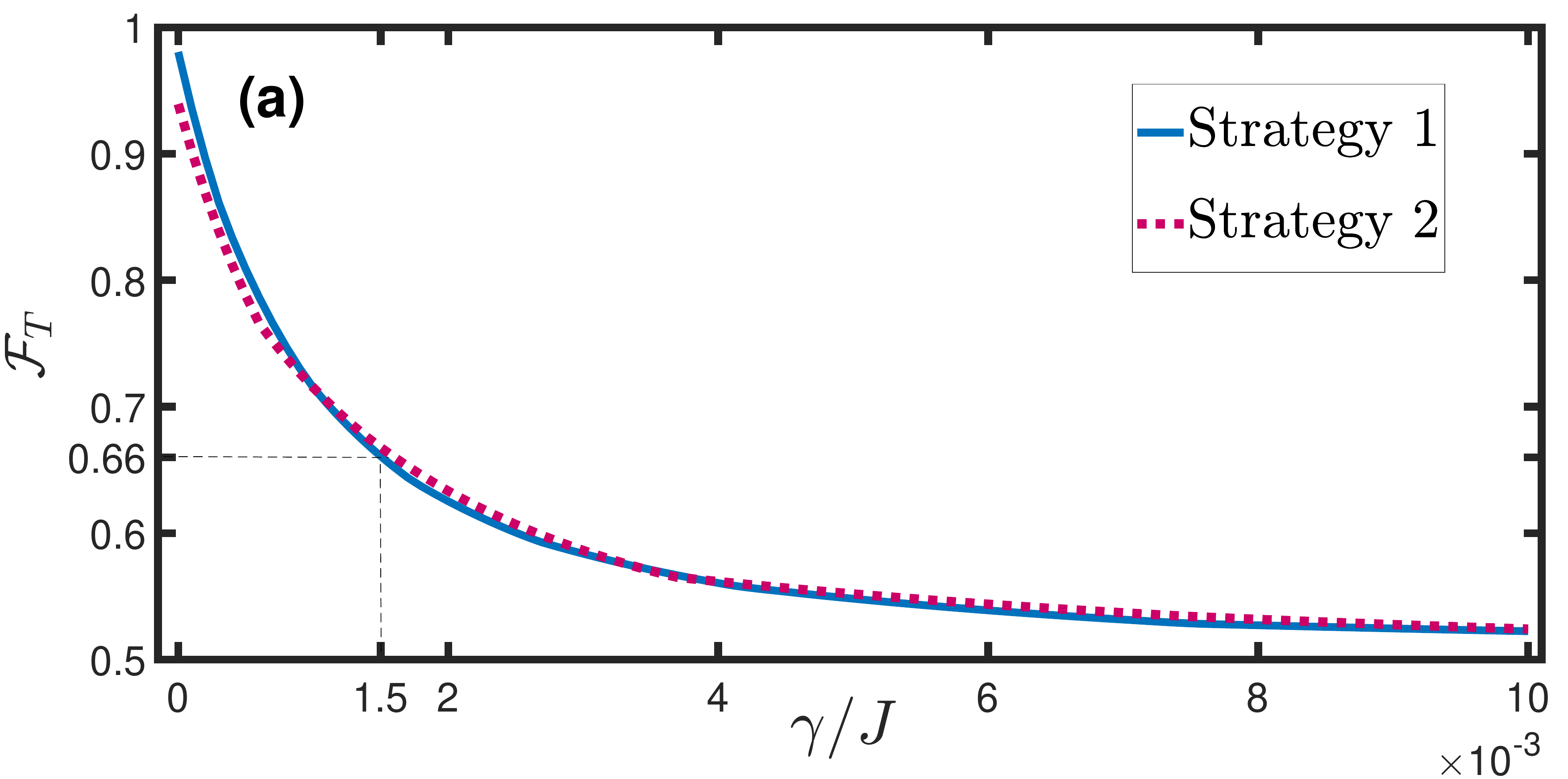} 
\includegraphics[width=\linewidth,height=3.8cm]{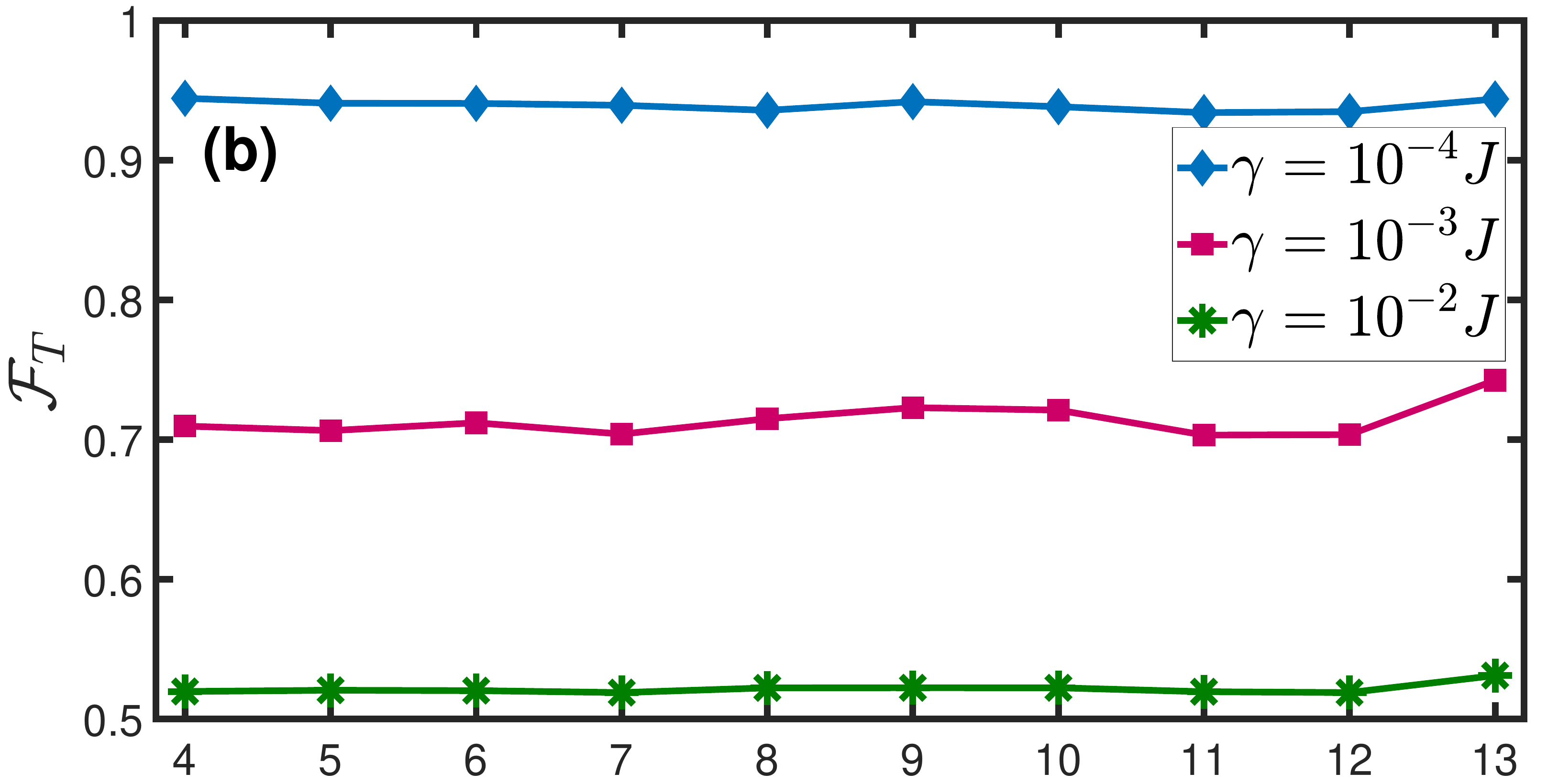}
\includegraphics[width=\linewidth,height=4cm]{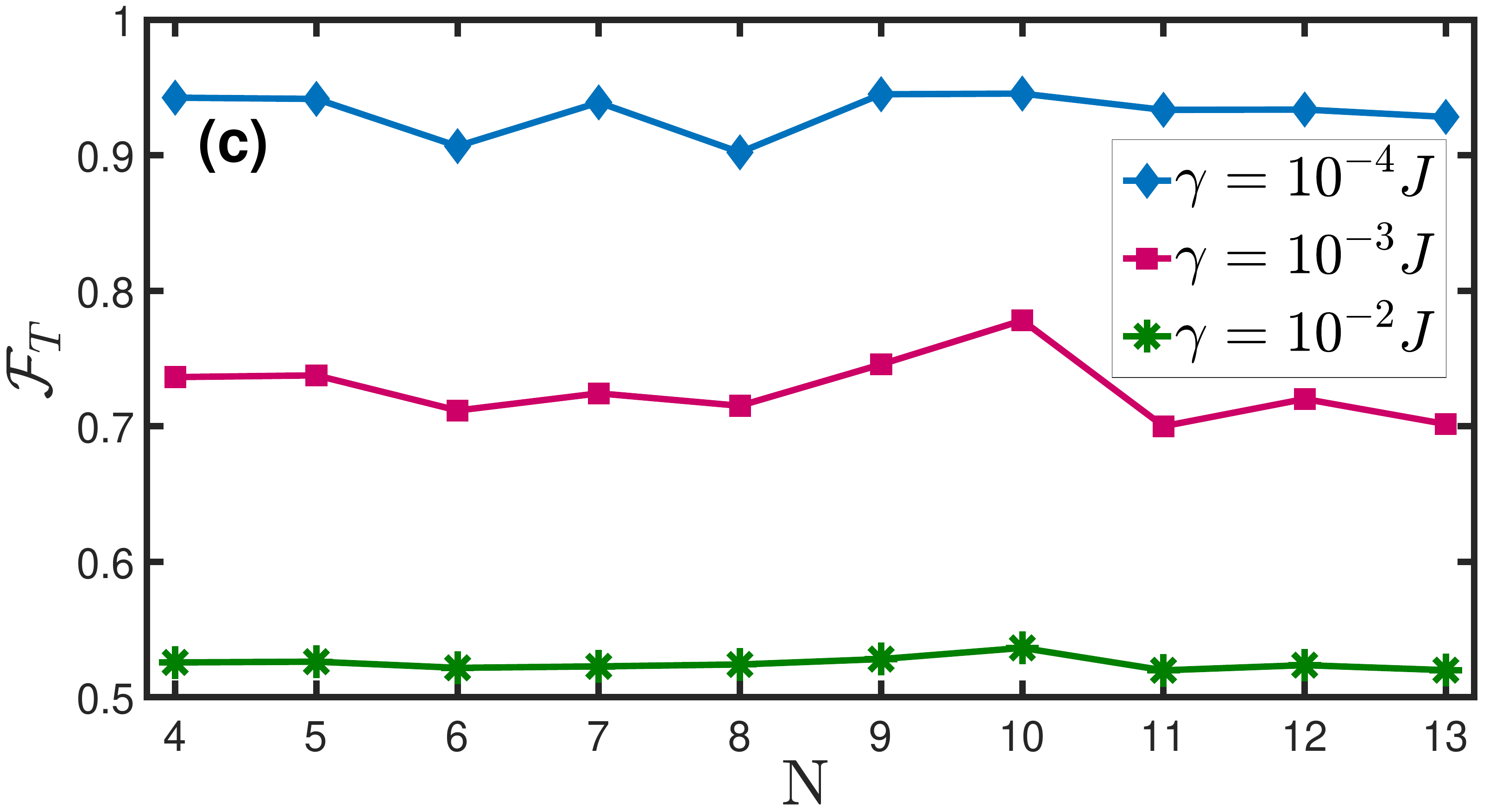}
\caption{Dephasing: (a) The average transmission fidelity for strategies $1$ and $2$ as function of dephasing rate $\gamma/J$ in a spin chain of length $N{=}8$. The Hamiltonian parameters are taken as $\{J_0/J{=}0.04, B_1/J{=}0.15, B_2/J{=}{-}0.05\}$ and $\{B_0/J{=}26, B_1/J{=}0.3, B_2/J{=}{-}0.2\}$, respectively, for strategies $1$ and $2$. (b) and (c) The average transmission fidelity as function of $N$ for three values of $\gamma/J$ and strategies $1$ and $2$, respectively. In preparing these plots the Hamiltonian parameters are tuned such that if $\gamma{=}0$, $\mathcal{F}_T$s  are equal with their maximum values for relevant strategy and $N$. }\label{fig:dephsing}
\end{figure}

A central aspect of all quantum processes in a real-world scenario is dephasing. It destroys the coherent superposition of quantum states and results in a classical mixture.
In a typical quantum state transfer protocol, the channel and users are not well isolated from the environment and might be disturbed by the effect of surrounding fluctuating magnetic or electric fields. This yields to random level fluctuations in the system and affects the fidelity of transmission.
For fast and weak random field fluctuations, i.e. in the Markovian limit, the evolution of the system can be described by a quantum master equation as 
\begin{eqnarray}\label{master}
\dfrac{d\rho(t)}{dt}=-i[H,\rho(t)]+ \gamma\sum_{i} \big(\sigma_{i}^{z}\rho(t)\sigma_{i}^{z} - \rho(t)\big),
\end{eqnarray}
where the first term in the right-hand side is the unitary evolution of the system and the second term is the
dephasing which acts on all the qubits involved with the rate $\gamma$. 
The fidelities $\overline{F}_{\alpha\beta}(t)$ (with $\alpha,\beta=1,\cdots ,M$) can be computed using the Eq.~(\ref{eq:Average Fidelity for 2 users}) in Appendix A.
To see the destructive effect of dephasing, in Fig.~\ref{fig:dephsing}(a), we plot the maximum average transmission fidelity as a function of $\gamma$ in a chain of length $N{=}8$ with $2$ users for our first two strategies. Clearly, by increasing the strength of the noise the quality of transmission decreases for both strategies. Nonetheless, for the dephasing rate $\gamma<1.5{\times} 10^{-3} J$ the fidelity remains above the classical threshold $2/3$.   
To finalize our analysis, in Figs.~\ref{fig:dephsing}(b) and (c), we depict the fidelity as a function of length $N$ for three values of $\gamma$, using strategy $1$ and $2$ respectively.  
The results show that the obtainable transmission fidelity only changes by the value of $\gamma$ and not the system size $N$. 
This is because the channel qubits are hardly populated and thus Lindbladian terms acting on channel qubits hardly change the state of the system.
It is worth mentioning that the slight fluctuations in the maximum values of the average transmission fidelity is because of the dependency of $\tau$ on the system size $N$. 

\section{Experimental proposal}
The best physical platform to provide the XX Hamiltonian with the required controllability of our protocol is superconducting coupled qubits. Recently, they have been used for simulating non-equilibrium dynamics of many-body systems for single-user perfect state transfer~\cite{li2018perfect}, many-body localization~\cite{xu2018emulating,chiaro2019growth}, spectrometry~\cite{roushan2017spectroscopic} and quantum random walks~\cite{yan2019strongly}. In such devices, the exchange coupling varies between $J {\sim} 10-50$ MHz, the dephasing time is $T_2=\sim 10-20$ $\mu$s (i.e. $\gamma=50$ KHz) and the local energy splitting, equivalent to magnetic fields in our protocol, can be tuned up to $800$ MHz (namely $B/J{\sim} 16$)~\cite{li2018perfect,xu2018emulating,chiaro2019growth,roushan2017spectroscopic,yan2019strongly}. Adopting our strategy 2, for a system of length  $N{=}8$ and exchange coupling $J{=}50$~MHz, one can tune the energy splittings to be $B_0{=}600$ MHz (i.e. $B_0/J{=}12$), $B_1{=}50$~MHz (i.e. $B_1/J{=}1$) and $B_2{=}{-}35$~MHz (i.e. $B_2/J{=}{-}0.7$). These parameters result in $\mathcal{F}_T{>}0.96$ for optimal time of $\tau{\simeq} 2.6$~$\mu$s, in the absence of decoherence. Considering the dephasing rate $\gamma/J=10^{-3}$ the fidelity  $\mathcal{F}_T$ is estimated to be $\sim 0.75$ which is still above $2/3$.


\section{Conclusion}
Spin chains provide fast and high-quality data buses for
connecting different registers. However, in the absence of simultaneous
communication between different sender-receiver
pairs, the speed of computation will be ultimately limited by
the waiting time required for the sequential use of the channel.
In this article, we address this key issue by introducing a
protocol for simultaneous quantum communication between
multiple users sharing a common spin chain data bus. Our proposal,
presented in three different strategies, is based on creating
an effective end-to-end interaction between each senderreceiver
pair and yields very high transmission fidelities. In
each proposed strategy, different sets of local parameters are
optimized so that each pair of users communicate through a
different energy eigenstate of the system. Since the energy
of each communication channel is off-resonance with the
others, the crosstalk is negligible.While all the three strategies
provide high transmission fidelities, the third strategy, which
is a hybrid control of both the coupling and the magnetic field,
outperforms the others. Moreover, increasing the number of
users does not significantly change the transmission time,
which means that the rate of communication is enhanced
proportional to the number of users. Our protocol is shown to be stable against various imperfections and can also be
realized in current superconducting quantum simulators.

\section{Acknowledgments}  

Discussions with Davit Aghamalyan, Kishor Bharti, and Marc-Antoine Lemonde are warmly acknowledged. A.B. acknowledges support from the National Key R$\&$D Program of
China, Grant No. 2018YFA0306703.


\appendix 
\addtocontents{toc}{\protect\contentsline{chapter}{Appendix:}{}}
\renewcommand\thefigure{A\arabic{figure}}
\setcounter{figure}{0}

\section{Average Fidelity Matrix for Excitation Conserving Hamiltonian}
The key mathematical objects needed to analyze the performance of simultaneous multiple-users quantum communication are 
$\overline{F}_{\alpha\beta}(t)$ evaluated by integration over the Bloch sphere of all possible pure input states.
To obtain a general form of these quantities, lets start by rewriting the Eq.~(\ref{eq:Initial state}) in the main text as 
\begin{eqnarray}\label{eq:Initial state App}
\vert \Psi_{0} \rangle &=& \sum_{\bm{i}} a_{\bm{i}}(\Theta)\vert \bm{i} \rangle\otimes \vert \bm{0}_{ch} \rangle
\otimes \vert \bm{0}_{R}\rangle 
\end{eqnarray}
where the vectors
$\vert \bm{i} \rangle=\vert i_{1},\cdots ,i_{M} \rangle$ ($i_{\alpha}=0,1$), $\vert\bm{0}_{ch}\rangle=\vert 0,\cdots ,0\rangle$ and $\vert\bm{0}_{R}\rangle=\vert 0,\cdots ,0\rangle$ denote the state of the senders, channel and receivers, respectively.
The coefficient $a_{\bm{i}}(\Theta)$ is an abbreviation for $a_{\bm{i}}(\Theta)=a_{i_{1},\cdots ,i_{M}}(\Theta)$ and contains all the parameters $\Theta=\{\theta_1,\cdots ,\theta_M,\phi_1,\cdots , \phi_M \}$ which are inputted by the $M$ senders.    
Considering the evolved state of the overall system as 
$\rho(t)=\mathcal{U}[\vert \Psi_{0} \rangle\langle \Psi_{0} \vert]$,  with $\mathcal{U}[\bullet]=e^{-iHt}\bullet e^{iHt}$ 
the output state of each receiver can be obtained by tracing out the other qubits as
\begin{eqnarray}\label{eq:Receiver_state}
\rho_{_{R_{\alpha}}}(t)=\sum_{\bm{i},\bm{j}} a_{\bm{i}}(\Theta)a_{\bm{j}}^{*}(\Theta)
\Gamma_{\bm{i},\bm{j}}^{\alpha}(t), \qquad  \alpha=1,\cdots ,M
\end{eqnarray} 
where 
$\Gamma_{\bm{i},\bm{j}}^{\alpha}(t)=Tr_{\widehat{R_{\alpha}}}(\mathcal{U}[\vert \bm{i} \rangle\langle \bm{j} \vert \otimes \vert \bm{0}_{ch} \rangle\langle \bm{0}_{ch} \vert \otimes \vert \bm{0}_{R} \rangle\langle \bm{0}_{R} \vert])$ and $Tr_{\widehat{R_\alpha}}$ means tracing over all sites except the receiver $R_\alpha$.
Substituting Eq.~(\ref{eq:Receiver_state}) in the fidelity $F_{\alpha\beta}(t,\Theta)=\langle \psi_{_{S_\alpha}}\vert \rho_{_{R_\beta}}(t) \vert \psi_{_{S_\alpha}} \rangle$ ($\alpha,\beta=1,\cdots ,M$) and taking the average over all
possible initial states on the surface of the Bloch spheres for all users, results in
\begin{eqnarray}\label{eq:Average Fidelity}
\overline{F}_{\alpha\beta}(t)&=&\int F_{\alpha\beta}(t,\Theta) d\Omega_{1} \cdots  d\Omega_{M} 
 \cr
&=&\frac{1}{2}
+\frac{1}{3\times 2^{M}}\Big \lbrace 
\sum_{\bm{i}\mid_{i_{\alpha}=0}}\langle 0 \vert\Gamma_{\bm{i},\bm{i}}^{\beta}\vert 0 \rangle
-\sum_{\bm{i}\mid_{i_{\alpha}=1}}\langle 0 \vert\Gamma_{\bm{i},\bm{i}}^{\beta}\vert 0 \rangle
\cr &+& \sum_{\bm{i,i'}\mid_{i_{\alpha}\neq i'_{\alpha}}}\langle i_{\alpha} \vert\Gamma_{\bm{i},\bm{i'}}^{\beta}\vert i^{'}_{\alpha} \rangle
\Big\rbrace,
\end{eqnarray}
where the first and second summations contain all  $\vert\bm{i}\rangle$ in which $i_{\alpha}=0$ and $i_{\alpha}=1$, respectively.
While the last summation includes all $\vert\bm{i}\rangle$ and $\vert\bm{i'}\rangle$ that only in $i_{\alpha}$ are different.

For the sake of completeness, we present the form of $\overline{F}_{\alpha\beta}(t)$ for a protocol with two users (i.e. $M=2$).
In this case the vector $\vert \bm{i}\rangle$ belongs to $\lbrace\vert 00\rangle, \vert 01\rangle, \vert 10\rangle, \vert 11\rangle\rbrace$. So, using Eq.~(\ref{eq:Average Fidelity}) results in 
\begin{eqnarray}\label{eq:Average Fidelity for 2 users}
\overline{F}_{1\beta}(t)&=&
\frac{1}{2}
+\frac{1}{12}\Big \lbrace 
\langle 0 \vert\Gamma_{00,00}^{\beta}+\Gamma_{01,01}^{\beta}\vert 0 \rangle
-\langle 0 \vert\Gamma_{10,10}^{\beta}+\Gamma_{11,11}^{\beta}\vert 0 \rangle
\cr &+& \langle 0 \vert\Gamma_{00,10}^{\beta}+\Gamma_{01,11}^{\beta}\vert 1 \rangle
+\langle 1 \vert\Gamma_{10,00}^{\beta}+\Gamma_{11,01}^{\beta}\vert 0 \rangle
\Big\rbrace,
\nonumber \\
\overline{F}_{2\beta}(t)&=&
\frac{1}{2}
+\frac{1}{12}\Big \lbrace 
\langle 0 \vert\Gamma_{00,00}^{\beta}+\Gamma_{10,10}^{\beta}\vert 0 \rangle
-\langle 0 \vert\Gamma_{01,01}^{\beta}+\Gamma_{11,11}^{\beta}\vert 0 \rangle
\cr &+& \langle 0 \vert\Gamma_{00,01}^{\beta}+\Gamma_{10,11}^{\beta}\vert 1 \rangle
+\langle 1 \vert\Gamma_{01,00}^{\beta}+\Gamma_{11,10}^{\beta}\vert 0 \rangle
\Big\rbrace.
\end{eqnarray}


\section{State dependency }
\begin{figure}[t!]
\centering\offinterlineskip
\includegraphics[width=\linewidth]{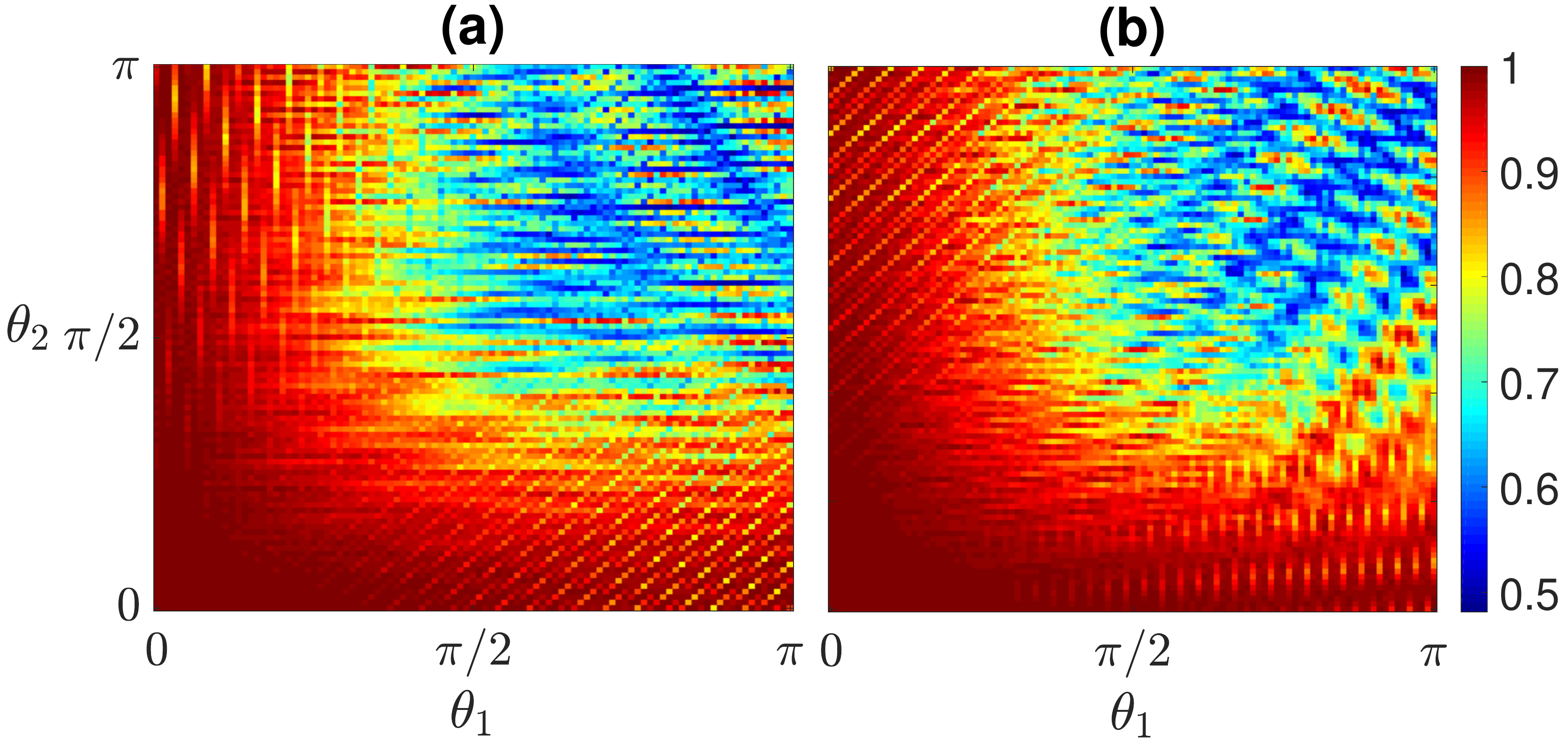} 
\caption{State dependency: The transmission fidelity for strategies $1$ (a), and $2$ (b) as function of polar angels $\theta_{1}$ and $\theta_{2}$ in a spin chain of length $N{=}20$.
Here azimuthal angels $\phi_{1}$ and $\phi_{2}$ are chosen as random numbers belongs to $[0,2\pi]$.
The Hamiltonian parameters for strategies $1$ and $2$ are taken as $\{J_{0}/J=0.04,B_1/J=0.35,B_2/J=-0.25\}$ and $\{B_{0}/J=21,B_1/J=0.3,B_2/J=-0.35\}$, respectively. All the plotted quantities are dimensionless.}\label{fig:State dependency}
\end{figure}

So far, we have averaged the fidelities over all possible inputs. However, some may argue that it is better to know the performance of the protocol in the worst scenario, namely the minimum obtainable fidelity. Note that this minimum fidelity may only happen for very special cases in the Bloch sphere and thus it is always good to study both the minimum and average fidelities together. In order to investigate the fidelity for different states, in Fig.~\ref{fig:State dependency}(a), we plot the transmission fidelity $\mathcal{F}_T{=}(\overline{F}_{11}+\overline{F}_{22})/2$ as a function of polar angles $\theta_{1}$ and $\theta_{2}$ (see Eq.~(\ref{eq:Initial state}) in the main text) in a chain of length $N{=}20$ by adopting strategy $1$. Here, azimuthal angels $\phi_{1}$ and $\phi_{2}$ ate chosen as random numbers within the interval $[0,2\pi]$. The same quantity for strategy $2$ is plotted in Fig.~\ref{fig:State dependency}(b).
As one can see, in both strategies, $\mathcal{F}_T$ takes its minimum when $\theta_{1},\theta_{2}\in[\pi/2,\pi]$, i.e. the two states are in the southern hemisphere of the Bloch sphere. This is due to the special choice of the state of the channel in which all qubits are initialized in $|0\rangle$, namely at the north pole of the Bloch sphere.  The figures clearly show that the fidelity $\mathcal{F}_T$ is mostly very high and only in some special states it takes lower values.  
In providing these plots, the Hamiltonian parameters are adjusted on their optimal values.   


\section{Relevant Eigenstates Localized at Boundaries }
\begin{figure*}[t!]
\begin{minipage}{1\columnwidth}
 \hfill
  \includegraphics[width=8.7cm,height=7cm]{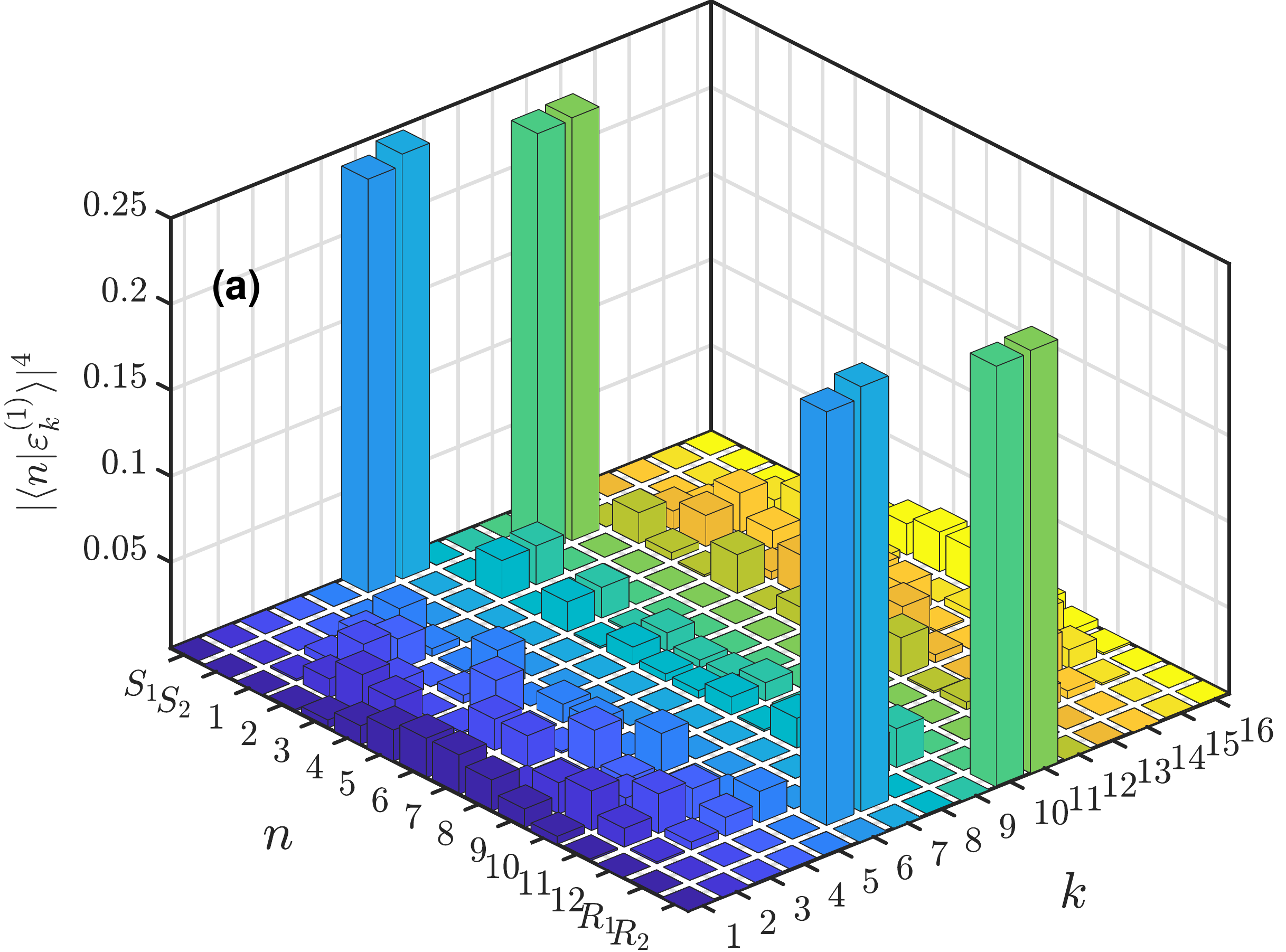}
\end{minipage}
\begin{minipage}{1\columnwidth}
\includegraphics[width=8cm,height=3.5cm]{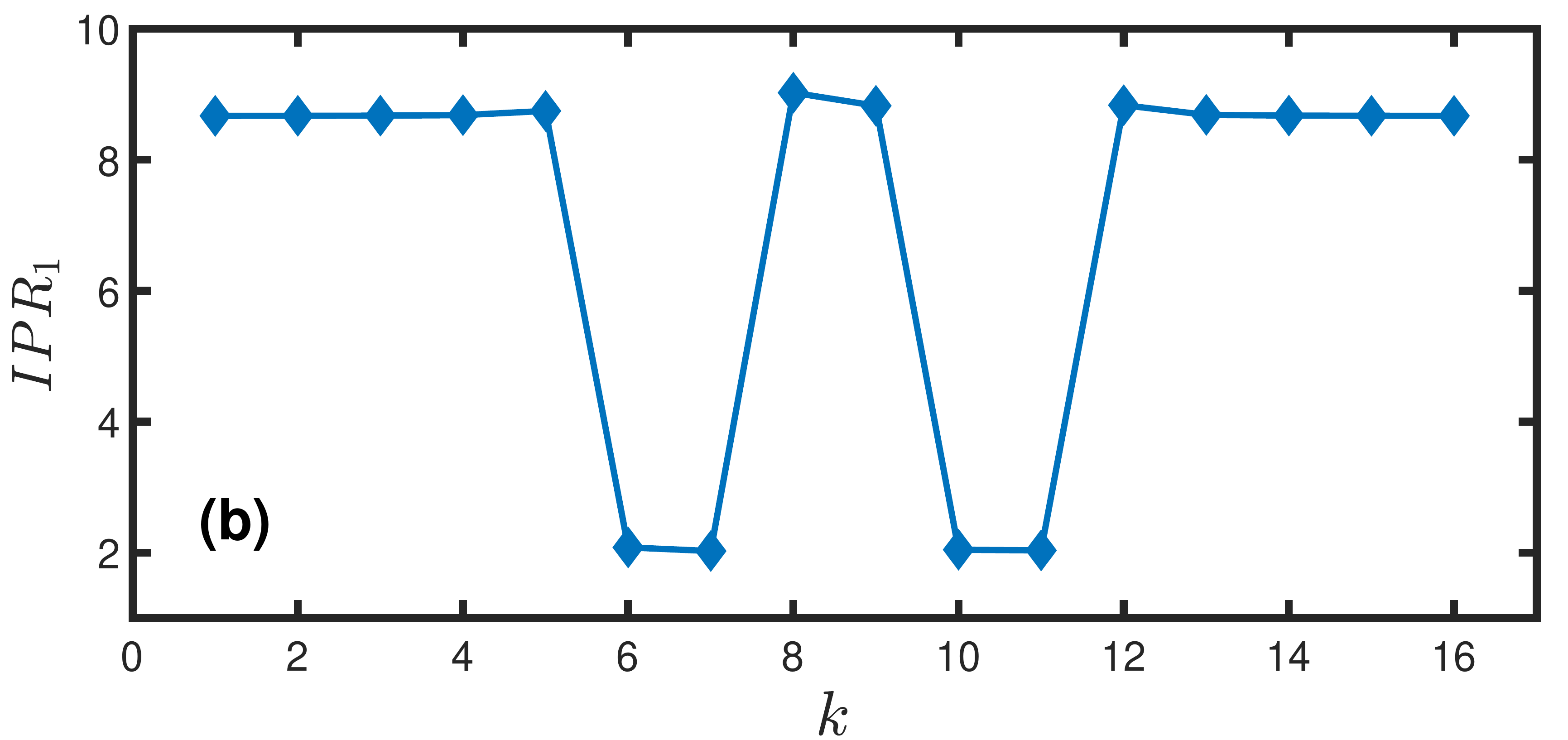}
 \hfill
\includegraphics[width=8cm,height=3.5cm]{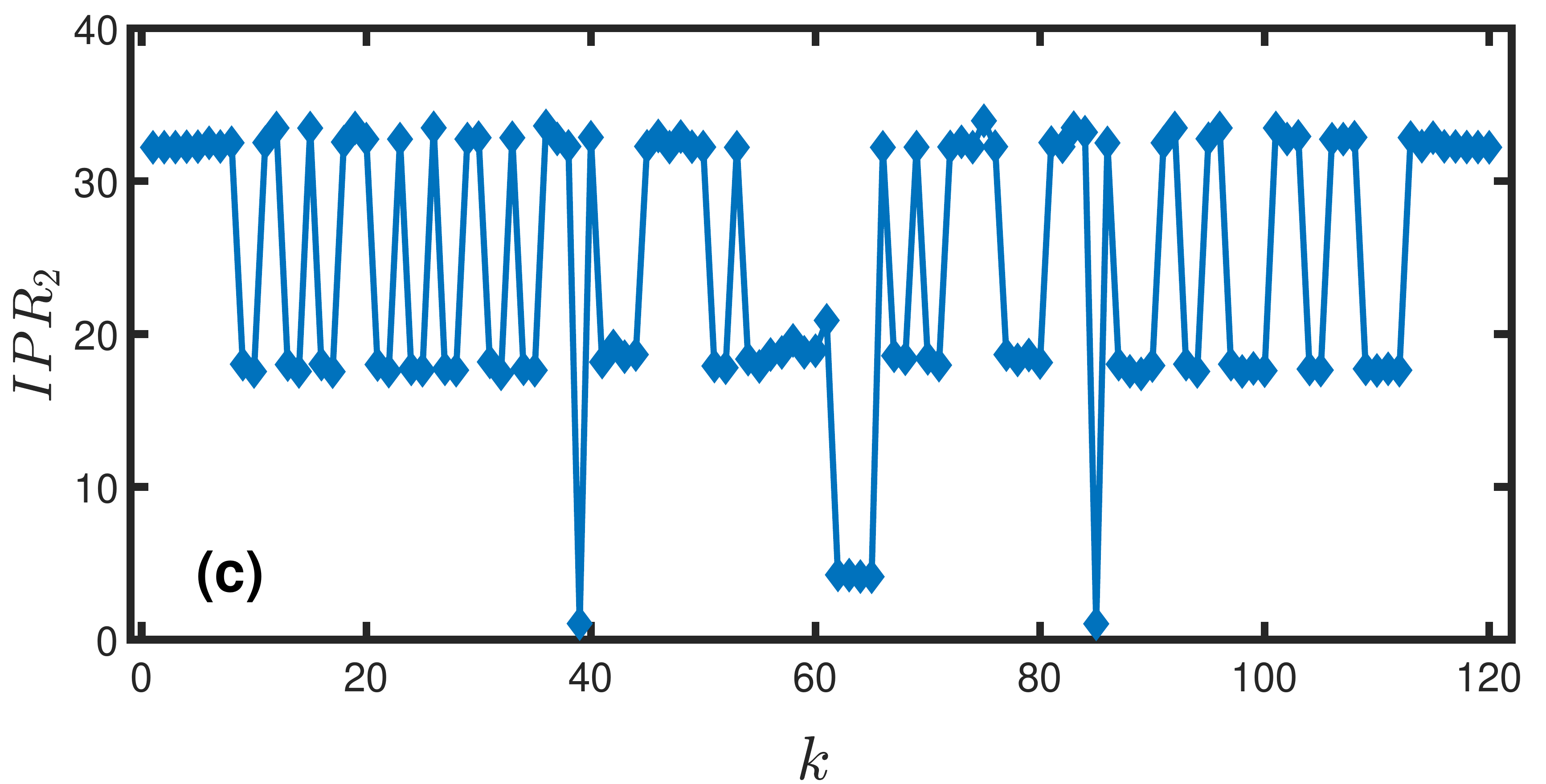}
\end{minipage}
\caption{Localization for $J_{0}\ll J$: (a) The localization value  $\vert\langle n \vert \varepsilon_{k}^{(1)} \rangle\vert^{4}$ of eigenstates with one-excitation in different position state $\vert n\rangle$. (b) The inverse
participation ratio $IPR_{1}$ as function of the number of  eigenstates in one-excitation subspace. (c) The inverse
participation ratio $IPR_{2}$ as function of the number of  eigenstates in two-excitation subspace. These quantities are obtained in chains of length $N{=}12$ with the Hamiltonian's parameters as $J_{0}/J{=}0.04$, $B_{0}/J{=}0$, $B_{1}/J{=}0.4$ and $B_{2}/J{=}{-}0.5$.}
\label{fig:IPR_Con1}
\end{figure*}

Making an effective end-to-end transmission between the senders and the receivers will be possible by either decreasing the coupling between users and the chain, i.e. choosing $J_0/J \ll 1$ or applying a strong magnetic field $B_0/J$ on the end sites of the chain.
In both cases, the excitations confine to the users' sites $\{S_1,\cdots ,S_M,R_1,\cdots ,R_M\}$ and leave the channel almost unexcited at all times.
Besides, by tuning the local magnetic fields $B_\alpha$, one can further localize   the excitations between each pair $(S_\alpha,R_\alpha)$ to minimize the crosstalk.
To investigate these issues we use the inverse
participation ratio (IPR), that will be defined below, to quantify the degree of localization of each Hamiltonian's eigenstate in different sites (e.g. see Ref.~\cite{lorenzo2013quantum}). 
Here, without loss of generality we restrict ourselves to the case of two users and particularly discuss the locality of eigenstates  in qubit sites $\{S_1,S_2,R_1,R_2\}$.    
Since XX Hamiltonian considered here commutes with the total spin in $z-$direction, and hence, conserves the number of excitations the dynamic of the overall system in the case of two users is restricted to evolve within the zero-, one- and two-excitation subspaces.
Let $\vert n \rangle$ ($ n\in \lbrace S_{1},S_{2},1,\cdots ,N,R_{1},R_{2}\rbrace$) and $\vert n_{1},n_{2} \rangle$ with $n_{1}<n_{2}$ ($n_{1}\in \lbrace S_{1},S_{2},1,\cdots ,N,R_{1}\rbrace$ and $n_{2}\in \lbrace S_{2},1,\cdots ,N,R_{1},R_{2}\rbrace$) denote the positions of the excitations in the one- and two-excitation subspaces, respectively.
Moreover, consider $\lbrace \varepsilon _{k}^{(\mu)}\rbrace$ and $\lbrace \vert \varepsilon _{k}^{(\mu)} \rangle\rbrace$ as the sets of the eigenvalues, in increasing order, and the corresponding 
eigenstates of $H^{(\mu)}$ ($\mu=1,2$) which in turn is the total Hamiltonian within the $\mu$-excitation subspace.
Since the type and the number of eigenstates of $H^{(1)}$ and $H^{(2)}$ are different, the IPR should be considered separately in each subspace.    
In one-excitation subspace the degree of localization of a given eigenstate $\vert \varepsilon _{k}^{(1)} \rangle$ can be calculated by $IPR_{1}$ defined as   
\begin{eqnarray}\label{IPR1}
IPR_{1}&=&\dfrac{1}{\sum_{n}\vert\langle n \vert \varepsilon_{k}^{(1)} \rangle\vert^{4}}.
\end{eqnarray}
When the eigenstate $\vert\varepsilon_{k}^{(1)}\rangle$ is highly localized, i.e. $\vert\langle n \vert \varepsilon_{k}^{(1)} \rangle\vert$ is nonzero for only one particular position state  $\vert n \rangle$, Eq.~(\ref{IPR1}) gets its minimum value, $1$, and when the eigenstate is uniformly distributed on all sites, this quantity attains its maximum value, $N$.

\begin{figure*}[t!]
\begin{minipage}{1\columnwidth}
 \hfill
  \includegraphics[width=8.8cm,height=7cm]{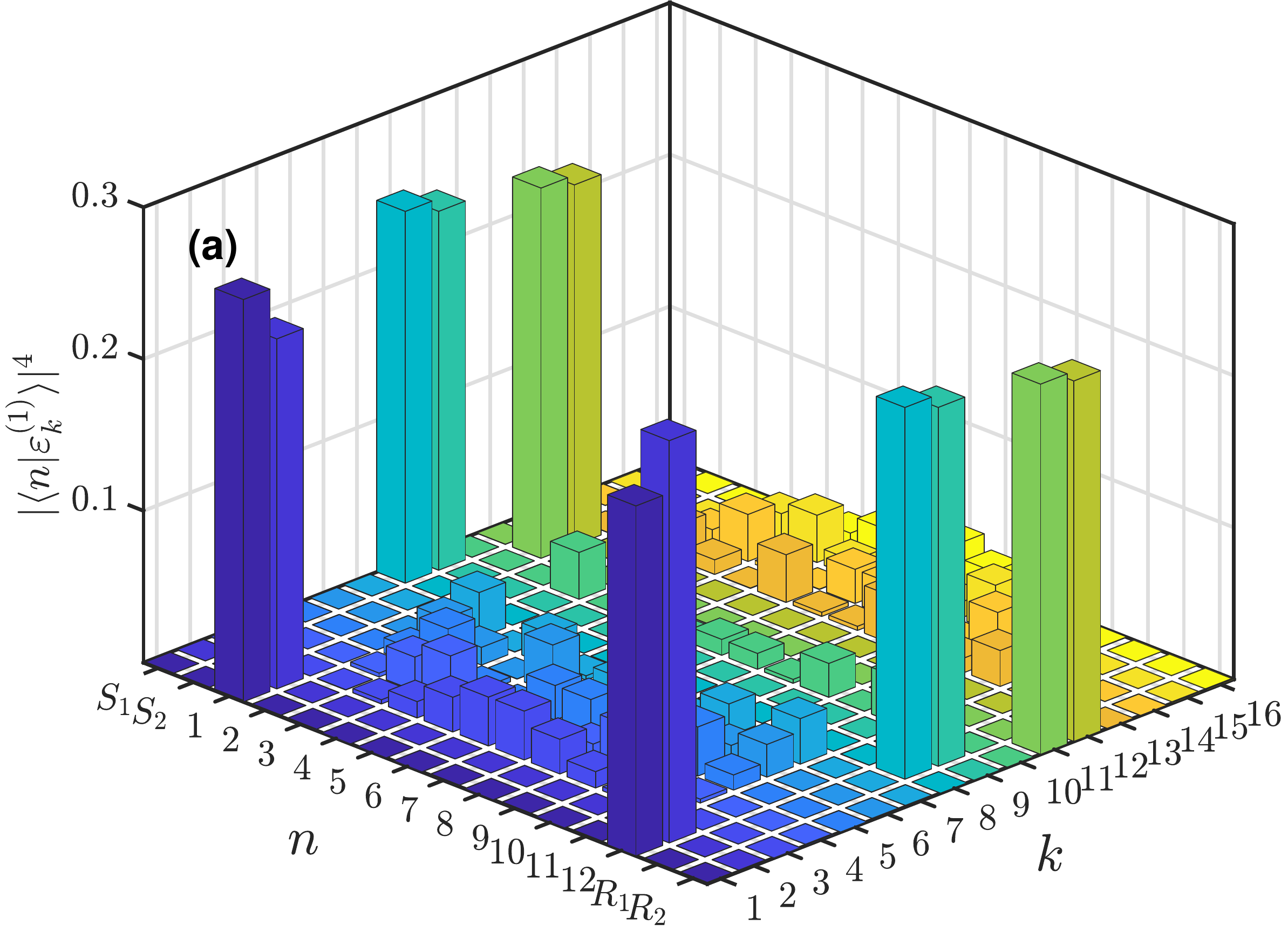}
\end{minipage}
\begin{minipage}{1\columnwidth}
\includegraphics[width=8cm,height=3.5cm]{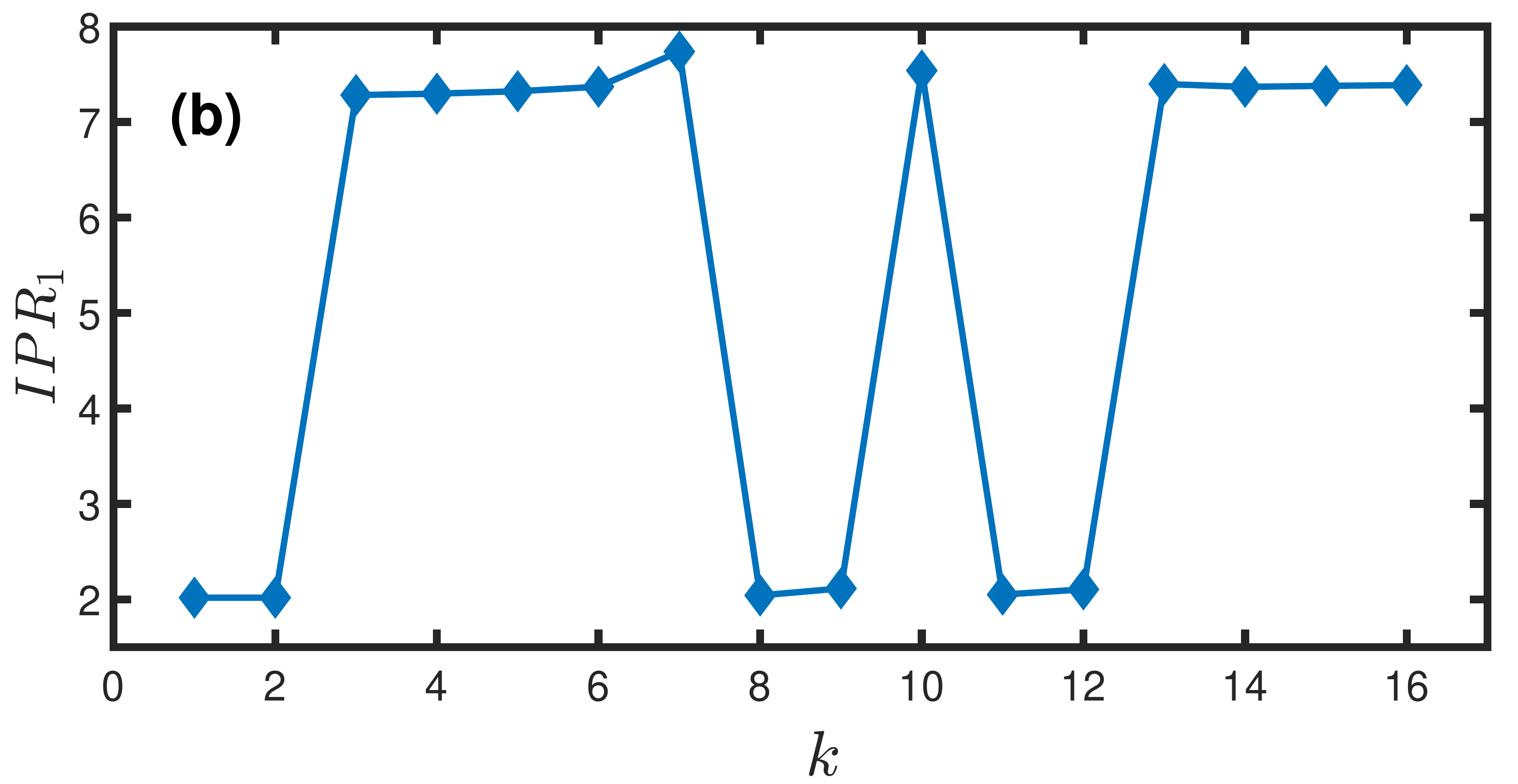}
 \hfill
\includegraphics[width=8cm,height=3.5cm]{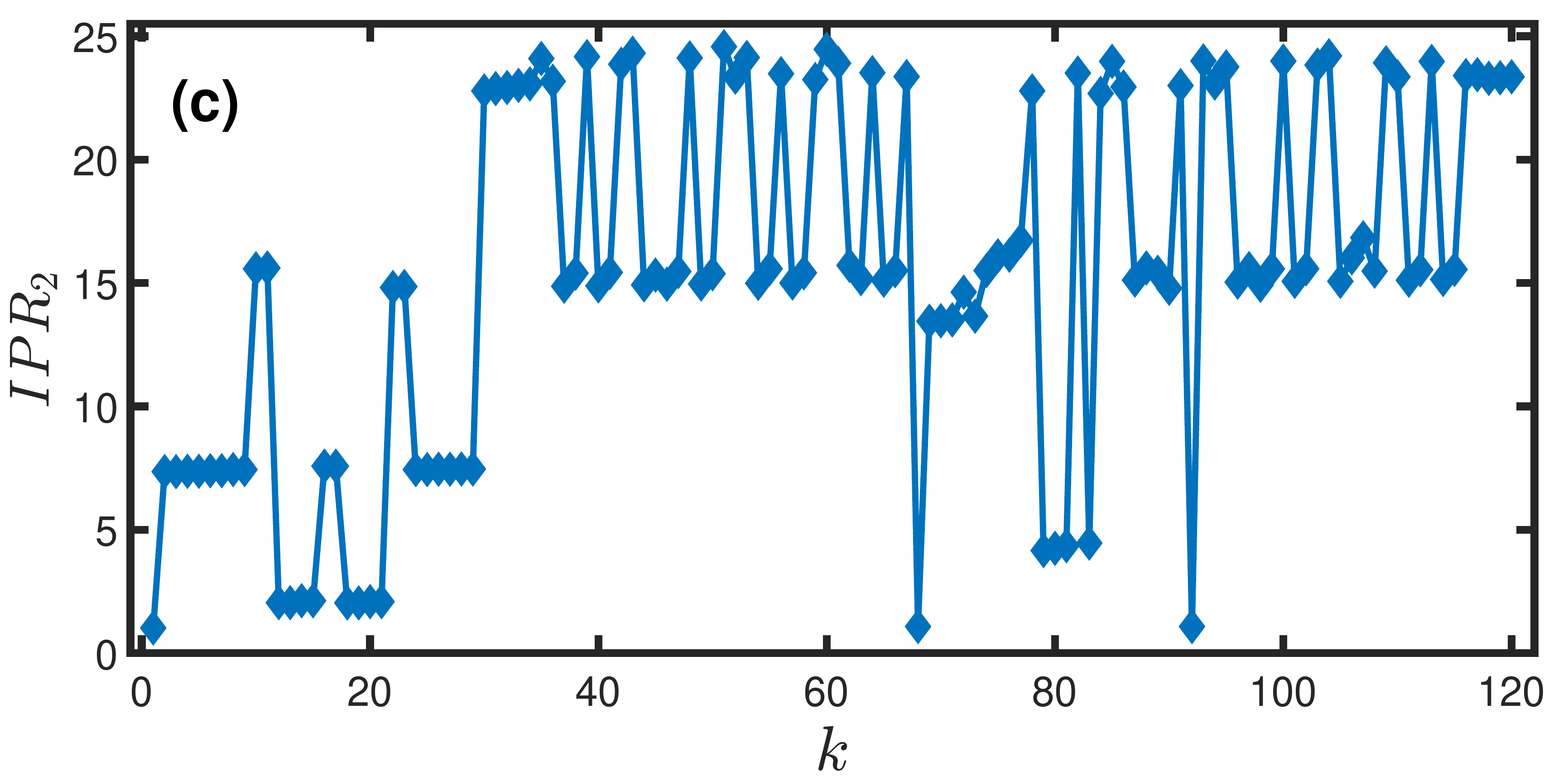}
\end{minipage}
\caption{Localization for $B_{0}>J$: (a) The localization value  $\vert\langle n \vert \varepsilon_{k}^{(1)} \rangle\vert^{4}$ of eigenstates with one excitation in different position state $\vert n\rangle$. (b) The inverse
participation ratio $IPR_{1}$ as  function of the number of eigenstates in one-excitation subspace. (c) The inverse
participation ratio $IPR_{2}$ as  function of the number of eigenstates in two-excitation subspace. These quantities are obtained in chains of length $N{=}12$ with the Hamiltonian's parameters which are considered as $J_{0}{/}J{=}1$, $B_{0}/J{=}25$, $B_{1}/J{=}0.15$ and $B_{2}/J{=}{-}0.45$.}\label{fig:IPR_Con2}
\end{figure*}

Likewise, for the eigenstates of $H^{(2)}$, the $IPR_{2}$ is
\begin{eqnarray}\label{IPR2} 
IPR_{2}&=&\dfrac{1}{\sum_{n_{1},n_{2}}\vert\langle n_{1},n_{2} \vert \varepsilon_{k}^{(2)} \rangle\vert^{4}}.
\end{eqnarray}
Analogs to the previous case the minimum value of $IPR_{2}$ is equal to $1$ which indicates that the eigenstate $\vert\varepsilon_{k}^{(2)}\rangle$ is completely localized in a specific position state $\vert n_{1},n_{2} \rangle$ and it's maximum value, $\mathcal{O}(N^2)$, appears when excitations are distributed on all sites uniformly. 
In the following, we exploit the $IPR_{1}$ and $IPR_{2}$ to peruse the localization of the Hamiltonian's eigenstates for the first and second strategies outlined in the main text.

The first strategy is based on weakly coupling the users to the chain (i.e. $J_0/J\ll 1$ and $B_0=0$).    
The degree of localization for Hamiltonian's eigenstates in one-excitation subspace 
is reported in Fig.~\ref{fig:IPR_Con1}(b) for a chain of length $N=12$. 
Clearly, two couples of degenerate eigenstates $\vert \varepsilon_{k}^{(1)} \rangle$  are highly localized with $IPR_{1}=2$. 
By considering the numerator of $IPR_{1}$, i.e. $\vert\langle n \vert \varepsilon_{k}^{(1)} \rangle\vert^{4}$ plotted in  Fig.~\ref{fig:IPR_Con1}(a) as a function of $n$ and $k$, one can find that the excitations of these eigenstates are strongly localized on sites $(S_{1},R_{1})$ and $(S_{2},R_{2})$.
Analogs results can be obtained for eigenstates with two excitations. In Fig.~\ref{fig:IPR_Con1}(c) the localization's degree $IPR_{2}$ is plotted as a function of $k$.  
Strong localization $IPR_{2}=1$ take places for two eigenstates at position states $\vert S_{1},R_{1}\rangle$ and $\vert S_{2},R_{2}\rangle$. 
Besides these two, there are four eigenstates with middle energies that show non-negotiable localization, i.e.  $IPR_{2}<5$. Our results show that these eigenstates have remarkable overlap only with $\vert S_{1},S_{2}\rangle, \vert S_{1},R_{2}\rangle, \vert S_{2},R_{1}\rangle$ and  $\vert R_{1},R_{2}\rangle$.  
Note that in producing Fig.~\ref{fig:IPR_Con1}, Hamiltonian's parameters are set as $B_1/J=0.4$, $B_2/J=-0.5$ and $J_0/J=0.04$.

Excitation confinement to the users' qubits can be also established by applying magnetic field $B_0$ on the end sites of the chain (corresponding to the second strategy outlined in the main text).
This is shown in Fig~\ref{fig:IPR_Con2}(b) for eigenstates in one-excitation subspace in a chain of length $N=12$.
In contrast to the previous case, the excitations are localized not only on users' sites but also on the first and last sites of the chain (see Fig~\ref{fig:IPR_Con2}(a)). 
Obviously, in the presence of $B_0$ two eigenstates corresponding to the lowest energy are extremely localized at the endest sites of the chain i.e. $(1, N)$.
It should be emphasized that these eigenstates can never be populated because of the barriers made by local magnetic field $B_0$.
The rest of the two couples eigenstates with $IPR_{1}{\simeq}2$ are high-localized in users' positions $\lbrace S_{1}, R_{1}\rbrace$ and $\lbrace S_{2}, R_{2}\rbrace$.
In Fig.~\ref{fig:IPR_Con2}(c), the localization of $\vert\varepsilon_{k}^{(2)}\rangle$ is also considered. 
Our results show that, while there are three eigenstates whit $IPR_{2}{=}1$ that completely overlap with three states $\vert 1,N\rangle$, $\vert S_{1},R_{1}\rangle$ and $\vert S_{2},R_{2}\rangle$, the others with remarkable localization (i.e. $IPR_{2}{<}5$) have superposition with the states belong to  
$\Big\{ \vert S_{1},1\rangle, \vert 1,R_{1}\rangle,\vert S_{1},N\rangle, \vert N,R_{1}\rangle, \vert S_{2},1\rangle, \vert 1,R_{2}\rangle, \vert S_{2},N\rangle, \vert N,R_{2}\rangle,$ $\vert S_{1},S_{2}\rangle, \vert S_{1},R_{2}\rangle, \vert S_{2},R_{1}\rangle, \vert R_{1},R_{2}\rangle \Big\}$.
Note that in Fig.~\ref{fig:IPR_Con2} the parameters of the Hamiltonian are tuned as $B_1/J=0.15$ $B_2/J=-0.45$, $B_0/J=25$, and $J_0/J=1$. \\ \\



\end{document}